\documentclass[number,preprint,3p]{elsarticle}

\usepackage{multirow}
\usepackage{color}
\usepackage{graphicx}
\usepackage{subcaption}
\usepackage{algorithm}
\usepackage{bm}
\usepackage[colorlinks]{hyperref} 
\usepackage{amssymb}
\usepackage{amsthm}
\usepackage{amsmath}
\usepackage{booktabs}
\usepackage{url}
\usepackage{scrextend}
\usepackage{epstopdf}
\usepackage{float}
\usepackage{tablefootnote}
\usepackage[table]{xcolor} 
\usepackage{mathtools}
\usepackage{soul}
\usepackage[perpage]{footmisc} 

\usepackage{lineno}

\makeatletter
\gdef\urlauthor#1#2{\g@addto@macro\@elsuads{\let\corref\@gobble%
     \def\@@tmp{#1}\raggedright\eadsep
     {\ttfamily\url{\expandafter\strip@prefix\meaning\@@tmp}}\space(#2)%
     \def\eadsep{\unskip,\space}}%
}
\gdef\emailauthor#1#2{\stepcounter{ead}%
     \g@addto@macro\@elseads{\raggedright%
      \let\corref\@gobble\def\@@tmp{#1}%
      \eadsep{\ttfamily\href{mailto:\expandafter\strip@prefix\meaning\@@tmp}{\expandafter\strip@prefix\meaning\@@tmp}}
      (#2)\def\eadsep{\unskip,\space}}%
}
\makeatother

\def\tr{^\intercal}

\newcommand{\vect}[1]{\boldsymbol{#1}}

\makeatletter
\let\@afterindenttrue\@afterindentfalse
\makeatother

\biboptions{numbers,sort&compress,square}
\journal{Structural Safety}
\begin{document}
\begin{frontmatter}
\renewcommand{\thefootnote}{\fnsymbol{footnote}}

\title{A composition of simplified physics-based model with neural operator for trajectory-level seismic response predictions of structural systems}

 \author[1]{Jungho Kim}
 \author[2]{Sang-ri Yi}
 \author[3]{Ziqi Wang\corref{cor1}}
\ead{ziqiwang@berkeley.edu}  
         \cortext[cor1]{Corresponding author}
 \address[1]{Department of Civil and Environmental Engineering, Sejong University, Seoul, South Korea}
 \address[2]{Department of Civil and Environmental Engineering, Rice University, United States} 
 \address[3]{Department of Civil and Environmental Engineering, University of California, Berkeley, United States}
\begin{abstract}
Accurate prediction of nonlinear structural responses is essential for earthquake risk assessment and management. While high-fidelity nonlinear time history analysis provides the most comprehensive and accurate representation of the responses, it becomes computationally prohibitive for complex structural system models and repeated simulations under varying ground motions. To address this challenge, we propose a composite learning framework that integrates simplified physics-based models with a Fourier neural operator to enable efficient and accurate trajectory-level seismic response prediction. In the proposed architecture, a simplified physics-based model, obtained from techniques such as linearization, modal reduction, or solver relaxation, serves as a preprocessing operator to generate structural response trajectories that capture coarse dynamic characteristics. A neural operator is then trained to correct the discrepancy between these initial approximations and the true nonlinear responses, allowing the composite model to capture hysteretic and path-dependent behaviors. Additionally, a linear regression-based postprocessing scheme is introduced to further refine predictions and quantify associated uncertainty with negligible additional computational effort. The proposed approach is validated on three representative structural systems subjected to synthetic or recorded ground motions. Results show that the proposed approach consistently improves prediction accuracy over baseline models, particularly in data-scarce regimes. These findings demonstrate the potential of physics-guided operator learning for reliable and data-efficient modeling of nonlinear structural seismic responses.
\end{abstract}

\end{frontmatter}
\renewcommand{\thefootnote}{\fnsymbol{footnote}}

\section{Introduction}
Accurate prediction of structural response trajectories under earthquake ground motion excitation is essential for a wide range of engineering applications, including structural design, maintenance planning, damage localization, and digital twin implementation~\cite{padgett2010risk,roy2013fundamental,lin2023digital,zhang2023seismic,jeon2025neural}. High-fidelity nonlinear time history analysis remains the primary computational approach for capturing the full complexity of structural dynamics. However, the high computational cost associated with repeated simulations under varying ground motions poses a significant challenge for real-time decision-making, probabilistic risk assessment, and parametric studies~\cite{moehle2004framework,kim2021quantile,morato2022optimal,kim2025uncertainty}.

To address this challenge, a variety of surrogate modeling techniques have been developed to emulate nonlinear structural responses at reduced computational cost. Classical approaches include Bayesian linear regression~\cite{ghosh2020seismic}, Kriging~\cite{kim2023estimation,yi2025multi}, polynomial chaos expansions~\cite{zhu2023seismic}, and deep neural networks~\cite{kim2020probabilistic,torky2021deep}. These models have shown promise in predicting selected engineering demand parameters but typically operate in finite-dimensional vector spaces and do not capture full response trajectories. To overcome this limitation, recent efforts have explored autoregressive formulations~\cite{schar2024feature,wang2024comparison}, state-space models~\cite{li2018gibbs,cheng2025state}, and deep sequence-to-sequence networks~\cite{du2018time,wang2020general} for time-history response prediction. For example, Zhang et al. (2019) \cite{zhang2019deep} applied long short-term memory (LSTM) networks to model temporal dependencies in nonlinear building systems, while Zhang et al. (2020) \cite{zhang2020physics} proposed a physics-guided convolutional neural network that incorporates physical constraints to improve prediction accuracy. More recently, He and Zhang (2025) \cite{he2025mc} developed a multi-channel gated recurrent unit-based model capable of generalizing nonlinear dynamics across varying structures and excitations. Recent works have also explored data-driven trajectory modeling using deep learning frameworks. For instance, Atila and Spence (2025) \cite{atila2025metamodeling} proposed a physics-informed LSTM model for nonlinear stochastic systems, while Li and Spence (2022) \cite{li2022metamodeling} introduced a deep surrogate framework for high-dimensional structural dynamics under stochastic excitation. Although these methods have shown promising results, these approaches often rely on restrictive assumptions such as narrow-band excitations or simplified system dynamics, limiting their applicability across diverse seismic scenarios. Notably, near-fault ground motions and ground motions on stiff soil sites often exhibit wide-band frequency characteristics, restricting the application of conventional dimensionality reduction-based surrogate modeling approaches. 

Neural operators offer a promising alternative by learning mappings between infinite-dimensional function spaces rather than finite-dimensional vectors~\cite{lu2021learning,li2020fourier,kovachki2023neural,kontolati2024learning}. Designed to approximate nonlinear solution operators of differential equations, i.e., a generalization of the Green's function, neural operators enable direct regression from input functions (e.g., ground motions) to output functions (e.g., structural response trajectories). Among existing formulations, the Fourier Neural Operator (FNO)~\cite{li2020fourier} introduces a spectral convolution framework that enables efficient, resolution-invariant learning. Other variants include Deep Operator Networks (DeepONet)~\cite{lu2021learning}, which separate functional and spatial encodings, and Graph-based Neural Operators (GNO)~\cite{li2020neural}, which support irregular spatial domains. Despite their success in scientific computing, the application of neural operators to nonlinear structural dynamics remains in its early stages.

Recent studies have begun to explore this potential. Goswami et al.~\cite{goswami2025neural} proposed a hybrid DeepONet–FNO framework for modeling structural responses under seismic and wind excitations. Cao et al.~\cite{cao2024deep} employed a wavelet DeepONet to predict the dynamic response of floating offshore structures subjected to irregular wave loading. Tainpakdipat et al.~\cite{tainpakdipat2025fourier} utilized the FNO to accelerate large-scale earthquake rupture simulations, while Perrone et al.~\cite{perrone2025integrating} integrated neural operators with diffusion models to enhance spectral compatibility in synthetic ground motion generation. Nevertheless, most existing efforts adopt fully data-driven formulations without leveraging physical insights derived from structural models. Moreover, performance under diverse ground motion inputs remains largely unexplored.

To enhance generalization and reduce data dependence, physics-guided machine learning has gained increasing attention. Physics-informed neural networks or operators attempt to embed governing partial differential equation (PDE) constraints directly into the loss function as soft penalties~\cite{goswami2023physics,li2024physics,azizzadenesheli2024neural,eshaghi2025variational}. While effective in cases where PDE residuals are available and differentiable, such methods are difficult to train and apply to structural systems governed by complex hysteretic behavior, where explicit residual and gradient forms may be unavailable or intractable.

This study proposes a different strategy: embedding structural dynamics into the learning architecture through a composite operator framework. Specifically, the proposed approach introduces a simplified physics-based simulator/operator between the seismic input and the neural operator. This intermediate step generates a coarse-grained response trajectory using domain-informed approximations such as linearization, modal reduction, or relaxed solvers with coarse time discretization~\cite{elishakoff2017sixty,peherstorfer2018survey,kramer2019nonlinear,patsialis2020reduced,wang2024optimized}. The neural operator is then trained to correct the discrepancy between this approximation and the full nonlinear response, allowing it to focus on modeling complex dynamics, such as hysteresis and path dependence, that are not captured by the simplified model. This composite architecture preserves the function-to-function learning paradigm while reducing the complexity of the solution operator and improving data efficiency.

While conceptually related strategies have been explored in quantity-of-interest-based surrogate modeling settings~\cite{xian2024physics,kim2024dimensionality}, their integration within neural operator learning for full trajectory-level structural response prediction remains limited. The present study addresses this gap by introducing a Composite Physics–Neural Operator (C-PhysFNO) framework that unites coarse-grained physics-based approximations with FNO-based operator learning. In addition, a linear regression-based postprocessing step is proposed to further refine predictions and quantify predictive uncertainty. This refinement operates in closed form and incurs negligible computational cost. The proposed framework is validated on three structural systems subjected to both synthetic and recorded ground motions. Numerical results demonstrate that the C-PhysFNO consistently outperforms standard neural operator baselines in both accuracy and data efficiency, particularly in limited-data regimes.

This composite learning approach shares conceptual similarity with prior work that augments physics-based models using data-driven corrections \cite{huang2023fast}. However, it differs in both scope and architectural design. For example, Huang et al. (2023) \cite{huang2023fast} proposed a neural vector-enhanced numerical solver for accelerating time integration in low-dimensional differential equation systems. In contrast, our framework targets the trajectory-level prediction of nonlinear seismic responses using operator learning. Moreover, by incorporating structured physics-based approximations—such as equivalent linearization and modal reduction—as intermediate representations, the C-PhysFNO model mitigates spectral bias and enhances predictive fidelity.

The remainder of this paper is organized as follows. Section~\ref{Background} introduces the formulation of neural operators and presents the nonlinear seismic response modeling problem. Section~\ref{Proposed} describes the proposed composite modeling architecture, physics-based input encoding, and training strategy. Section~\ref{Postprocessing} outlines the linear regression-based postprocessing approach. Section~\ref{Examples} presents numerical investigations across multiple structural examples and diverse ground motion types. Section~\ref{Conclusion} concludes the paper with a summary and directions for future work.

\section{Background} \label{Background}

\subsection{Neural operator learning and Fourier formulation} \label{NO_FNO}

Neural operators aim to learn mappings between infinite-dimensional function spaces, enabling direct approximation of solution operators in systems governed by differential equations. Let \( a(\vect{x}) \in \mathcal{A} \) denote an input function (e.g., time-varying excitations) defined over a domain \(\Omega \subset \mathbb{R}^d\), and let \( u(\vect{x}) \in \mathcal{U} \) denote the corresponding output function (e.g., structural response trajectories). The objective is to approximate a nonlinear operator
\begin{equation} \label{Eq:operator1}
\mathcal{G}: \mathcal{A} \to \mathcal{U}\,,
\end{equation}
such that
\begin{equation} \label{Eq:operator2}
u(\vect{x}) = \mathcal{G}(a)(\vect{x})\,, \quad \vect{x} \in \Omega\,,
\end{equation}
where $\vect{x}$ is a spatial or temporal variable, and $\mathcal{A}$, $\mathcal{U}$ are infinite-dimensional function spaces, such as the $L^2$ space of square-integrable functions.

In practice, the operator $\mathcal{G}$ is learned from a dataset $\{a_i(\vect{x}), u_i(\vect{x})\}_{i=1}^{N}$ consisting of input–output function pairs generated from simulations or experiments. A neural operator $\mathcal{G}_{\vect\theta}^{\mathrm{NO}}$ is typically constructed through a series of transformations on intermediate feature representations $v^{(l)}(\vect{x})$, governed by trainable parameters $\vect\theta$. A general formulation is given by
\begin{align}
v^{(0)}(\vect{x}) &= \mathcal{Q}(a(\vect{x}))\,, \quad &&\text{(Lifting layer)}, \label{Eq:NO_lift} \\
v^{(l+1)}(\vect{x}) &= \sigma\left( \mathcal{K}_l(v^{(l)})(\vect{x}) + \mathcal{W}_l(v^{(l)}(\vect{x})) \right)\,, \quad l=0,\dots,L-1, \quad &&\text{(Iterative kernel layers)} \label{Eq:NO_kernel} \\
u(\vect{x}) &= \mathcal{P}(v^{(L)}(\vect{x}))\,, \quad &&\text{(Projection layer)}, \label{Eq:NO_proj}
\end{align}
where $\mathcal{Q}$ is a lifting operator that embeds the input function into a higher-dimensional latent space, $\sigma(\cdot)$ is a nonlinear activation function (e.g., ReLU), $\mathcal{K}_l$ is a global kernel operator, typically an integral or convolution, $\mathcal{W}_l$ is a local transformation, and $\mathcal{P}$ projects the output back to the physical space. The complete operator is expressed as
\begin{equation} \label{Eq:NO_full} \mathcal{G}_{\vect\theta}^{\text{NO}} = \mathcal{P} \circ \sigma\left(\mathcal{K}_L + \mathcal{W}_L \right) \circ \cdots \circ \sigma\left(\mathcal{K}_1 + \mathcal{W}_1 \right) \circ \mathcal{Q}\,,
\end{equation}
where $\circ$ denotes function composition. Neural operator models typically accept structured input in the form of tensors arranged as ``Batch–Spatial–Channel," where the batch axis corresponds to multiple input samples, the spatial axis represents temporal or spatial discretization points, and the channel axis encodes features at each location. In this setting, the operator input is commonly expressed as $[a(\vect{x}), \vect{x}]$, indicating that at each location $\vect{x}$, the input function has value $a(\vect{x})$. The choice of kernel operator $\mathcal{K}_l$ defines different neural operator variants, including the FNO~\cite{li2020fourier}, DeepONet~\cite{lu2021learning}, and GNO~\cite{li2020neural}.


The FNO replaces the kernel integrals $\mathcal{K}_l$ with convolutions defined in the Fourier domain~\cite{li2020fourier}. This formulation enables the model to capture long-range dependencies efficiently and allows for resolution-invariant learning.

\begin{figure}[H]
  \centering
  \includegraphics[scale=0.55] {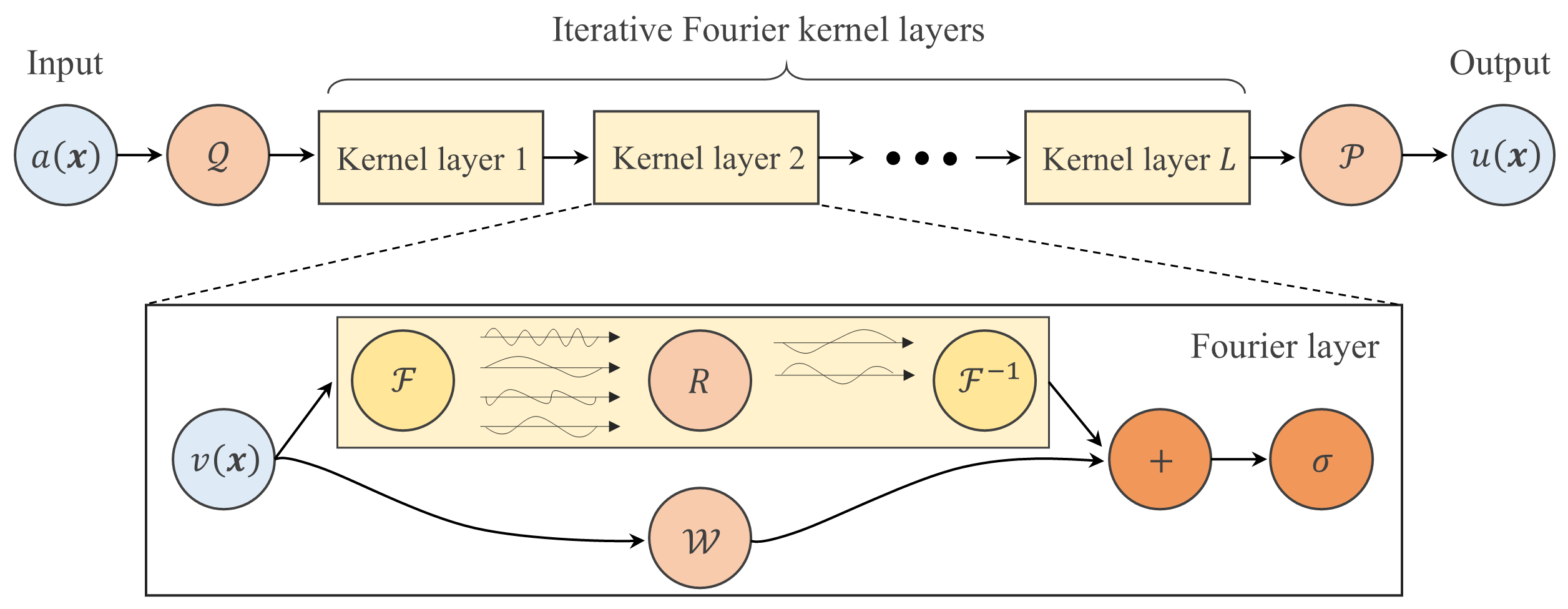}
  \caption{\textbf{Schematic of the FNO architecture~\cite{li2020fourier}.} The input function is lifted into a high-dimensional feature space, transformed through multiple spectral convolution layers, and projected back to the output space.}
  \label{Fig_FNO}
\end{figure}

At each layer $l$, FNO transforms the feature field $v^{(l)}(\vect{x})$ through a truncated Fourier series. The update rule for the FNO layer is
\begin{equation} \label{Eq:FNO_kernel}
v^{(l+1)}(\vect{x}) = \sigma\left( \mathcal{F}^{-1}\left( R^{(l)}(\mathcal{F}(v^{(l)})) \right)(\vect{x}) + W_{\mathrm{loc}}^{(l)}(v^{(l)}(\vect{x})) \right),
\end{equation}
where $\mathcal{F}$ and $\mathcal{F}^{-1}$ denote the forward and inverse Fourier transforms, $R^{(l)}$ is a learnable spectral multiplier, and $W_{\mathrm{loc}}^{(l)}$ is a local $1 \times 1$ convolution operator. The spectral multiplier modifies only the first $K$ modes to enforce sparsity, expressed as
\begin{equation} \label{Eq:FNO_filter}
R^{(l)}(\hat{v}^{(l)})(k) = 
\begin{cases}
W_{\mathrm{spec}}^{(l)}(k) \, \hat{v}^{(l)}(k), & \text{for} \,\,\, |k| \leq K, \\
0, & \text{otherwise},
\end{cases}
\end{equation}
where $\hat{v}^{(l)}$ is the Fourier transform of $v^{(l)}$, $W_{\mathrm{spec}}^{(l)}(k)$ are trainable complex-valued spectral weights, and $K$ is the truncation threshold. Thereby, the learnable parameters in each layer consist of the local weights $W_{\mathrm{loc}}^{(l)}$ and the spectral multipliers $W_{\mathrm{spec}}^{(l)}(k)$ for each retained mode. A schematic of the FNO architecture is illustrated in Figure~\ref{Fig_FNO}.

\subsection{Nonlinear seismic response modeling} \label{Problem}
Response of a nonlinear structural system under seismic excitation is governed by the equation of motion
\begin{equation} \label{Eq:problem}
\mathbf{M} \ddot{\vect{u}}(t) + \mathbf{C} \dot{\vect{u}}(t) + \mathbf{f}_{\mathrm{NL}}(\vect{u}(t), \dot{\vect{u}}(t)) = -\mathbf{M} \mathbf{1}a_g(t),
\end{equation}
where $\mathbf{M}$, $\mathbf{C} \in \mathbb{R}^{n_d \times n_d}$ denote the mass and damping matrices, respectively, $n_d$ is the number of degrees of freedom, $\vect{u}(t), \dot{\vect{u}}(t), \ddot{\vect{u}}(t) \in \mathbb{R}^{n_d}$ are the displacement, velocity, and acceleration vectors at time $t$, respectively, and $\mathbf{f}_{\mathrm{NL}}$ is the nonlinear restoring force. The ground acceleration $a_g(t)$ represents the input excitation, which may be synthetically generated using ground motion models or obtained from recorded earthquake events. The vector $\mathbf{1} \in \mathbb{R}^{n_d}$ denotes a vector of ones, used to apply uniform ground acceleration to all degrees of freedom.

The objective of operator learning in this context is to approximate the functional mapping from the seismic excitation $a_g(t)$ to the time-dependent multivariate structural response $\vect{u}(t) = [u_1(t), \dots, u_{n_d}(t)]^{\tr}$. However, direct training of neural operators on this mapping is often challenged by the complexity of nonlinear structural behavior, hysteresis, and the variability of seismic input characteristics.

\section{Composite physics-neural operator (C-PhysFNO) for seismic response prediction} \label{Proposed}

\subsection{Overview of the proposed approach} \label{Proposed_overview}

To enhance data efficiency and generalization in modeling nonlinear seismic responses, we propose a composite neural operator learning approach, termed C-PhysFNO, which integrates domain-informed, physics-based approximations into the learning process. Instead of directly learning the mapping from ground acceleration $a_g(t)$ to the nonlinear structural response $\vect{u}(t)$, the proposed approach introduces an intermediate representation $\vect{z}(t)$—a surrogate trajectory obtained from a simplified model. This representation captures coarse structural behavior and serves as a physics-guided preprocessing step. The neural operator is then trained to correct this approximation by learning the residual dynamics, thereby enabling more efficient and accurate trajectory-level response prediction.

The overall structure of the composite operator is defined as:
\begin{equation} \label{Eq:Physics_FNO}
\begin{aligned}
&\hat{\vect{u}}(t) =  \mathcal{G}^{\mathrm{NO}}_{\vect\theta}(\vect z)(t) =  \left( \mathcal{G}^{\mathrm{NO}}_{\vect\theta} \circ \mathcal{H}_p \right)(a_g)(t), \\
&\mathcal{H}_p : a_g(t) \in \mathcal{A} \mapsto \vect{z}(t) \in \mathcal{Z}, \\
& \mathcal{G}^{\mathrm{NO}}_{\vect\theta} : \vect{z}(t) \in \mathcal{Z} \mapsto \vect{u}(t) \in \mathcal{U}, \quad t \in [0,\tau],
\end{aligned}
\end{equation}
where $\mathcal{H}_p$ is a physics-based operator that maps the ground motion input $a_g(t)$ into a coarse-grained surrogate trajectory $\vect{z}(t)$, $\mathcal{G}^{\mathrm{NO}}_{\vect\theta}$ is a neural operator trained to correct the approximation and recover the full nonlinear structural response $\vect{u}(t)$, and $\vect{\theta}$ denotes the trainable parameters of the neural operator, and $\tau$ is the total duration of the response time window considered for prediction. This composite approach preserves the function-to-function learning paradigm central to operator-based models while introducing physics-based layer into the learning process.

The physics-based operator $\mathcal{H}_p$ leverages structural domain knowledge to produce a coarse approximation of structural dynamics using simplified physical models, such as linearization, modal reduction, or relaxed solvers with coarse time discretization. These intermediate representations are designed to retain dominant structural behavior while reducing modeling complexity. Practical implementations of $\mathcal{H}_p$ using various coarse-graining strategies are detailed in Section~\ref{Proposed_Hp}. The neural operator, introduced in Eq.~\eqref{Eq:NO_full}, is subsequently trained to correct the approximation $\vect{z}(t)$ and recover the true nonlinear response $\vect{u}(t)$.

\begin{figure}[H]
  \centering
  \includegraphics[scale=0.455] {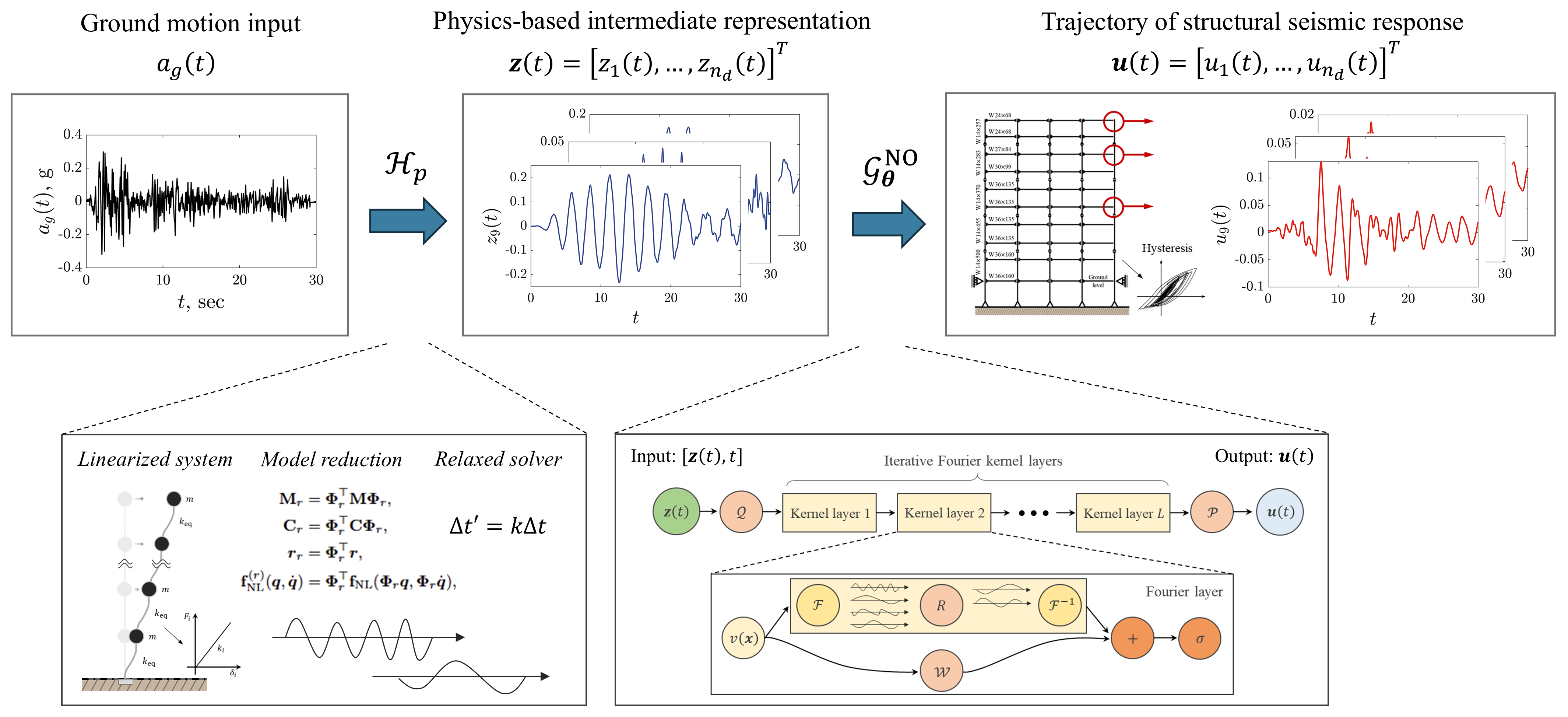}
  \caption{\textbf{Schematic of the proposed composite physics-neural operator framework.} Ground motion $a_g(t)$ is processed through a simplified physics-based simulator $\mathcal{H}_p$ to generate an intermediate trajectory $\vect{z}(t)$, which is then mapped to the nonlinear structural response $\vect{u}(t)$ by a neural operator $\mathcal{G}^{\mathrm{NO}}_{\vect{\theta}}$.}
  \label{Fig_illustration}
\end{figure}

Given a set of end-to-end paired trajectories $\{ a_g^{(i)}(t), \vect{u}^{(i)}(t) \}_{i=1}^{N}$ for training, the intermediate approximations $\vect{z}^{(i)}(t) = \mathcal{H}_p(a_g^{(i)}(t))$ are generated from the simplified physical model. Then, the neural operator is trained on $\{ \vect{z}^{(i)}(t), \vect{u}^{(i)}(t) \}_{i=1}^{N}$. An overview of the composite learning architecture is illustrated in Figure~\ref{Fig_illustration}.

It is noted that neural operators may exhibit spectral bias that favors the learning of low-frequency components over high-frequency ones \cite{rahaman2019spectral}. In seismic response prediction under broadband excitations, this bias can hinder accurate learning of complex dynamics. The proposed composite framework alleviates this issue by introducing a physics-based surrogate $\mathcal{H}_p$ that captures essential frequency content, thereby simplifying the residual function to be learned by the neural operator.

\subsection{Physics-based intermediate representations} \label{Proposed_Hp}

The intermediate operator $\mathcal{H}_p$ injects physical knowledge into the learning process by transforming unstructured input excitations into highly structured response trajectories. While there are different strategies for building a simplified physics model, three representative implementations of $\mathcal{H}_p$ are considered in this study: (i) equivalent linear system (behavioral simplification), (ii) modal reduction (spatial simplification), and (iii) relaxed solver (temporal simplification). Each offers a physically interpretable, lower-complexity surrogate that serves as the input to the neural operator.

\subsubsection{Equivalent linear system (ELS)} \label{Hp_ELS}

The ELS approach replaces the nonlinear structural model with a linear system characterized by effective stiffness matrix $\mathbf{K}_{\mathrm{eq}}$~\cite{elishakoff2017sixty}. The intermediate trajectory is obtained by solving:
\begin{equation} \label{Eq:ELS}
\mathbf{M} \ddot{\vect{z}}(t) + \mathbf{C} \dot{\vect{z}}(t) + \mathbf{K}_{\mathrm{eq}} \vect{z}(t) = -\mathbf{M} \mathbf{1} a_g(t),
\end{equation}
where $\mathbf{M}$ and $\mathbf{C}$ are the mass and damping matrices, and $\mathbf{K}_{\mathrm{eq}}$ can be determined from various linearization approaches~\cite{elishakoff2017sixty,wang2024optimized,kim2024dimensionality}. The most straightforward approach is to use the initial elastic stiffness of the structure. Refined approaches include minimizing the mean square error between the linear and nonlinear responses or maximizing the correlations between response quantities of interest. In this work, we adopt the simple approach of using the elastic stiffness. This is because: (i) we did not observe a noticeable performance gain by implementing a more complex ELS scheme in the C-PhysFNO pipeline; and (ii) seismic design codes typically require performing time history analysis on the linear elastic system before applying it to nonlinear systems. Therefore, from an engineering perspective, no additional modeling effort is required.

\subsubsection{Modal reduction} \label{Hp_modal}

Modal reduction provides a spatial coarse-graining technique by projecting the full-order model onto a reduced modal subspace. Let $\boldsymbol{\Phi}_r \in \mathbb{R}^{n_d \times r}$ denote the matrix of mode shapes corresponding to the $r$ lowest modes of the undamped linear system. The response is approximated as $\vect{z}(t) = \boldsymbol{\Phi}_r \vect{q}(t)$, where $\vect{q}(t) \in \mathbb{R}^r$ are reduced-order modal coordinates. Substituting this expression into the full nonlinear equation of motion and projecting the dynamics onto the reduced modal subspace yields the following reduced system:
\begin{equation} \label{Eq:modal_projected}
\mathbf{M}_r \ddot{\vect{q}}(t) + \mathbf{C}_r \dot{\vect{q}}(t) + \mathbf{f}_{\mathrm{NL}}^{(r)}(\vect{q}(t), \dot{\vect{q}}(t)) = -\mathbf{M}_r \vect{r}_r a_g(t),
\end{equation}
where the reduced matrices and internal force vector are defined as
\begin{align}
\mathbf{M}_r &= \boldsymbol{\Phi}_r^\top \mathbf{M} \boldsymbol{\Phi}_r, \\
\mathbf{C}_r &= \boldsymbol{\Phi}_r^\top \mathbf{C} \boldsymbol{\Phi}_r, \\
\vect{r}_r &= \boldsymbol{\Phi}_r^\top \vect{r}, \\
\mathbf{f}_{\mathrm{NL}}^{(r)}(\vect{q}, \dot{\vect{q}}) &= \boldsymbol{\Phi}_r^\top \mathbf{f}_{\mathrm{NL}}(\boldsymbol{\Phi}_r \vect{q}, \boldsymbol{\Phi}_r \dot{\vect{q}}),
\label{Eq:modal_projected2}
\end{align}
with $\vect{r}_r \in \mathbb{R}^{r}$ denoting the influence vector. This formulation preserves essential nonlinear behavior within a reduced number of degrees of freedom, enabling efficient simulation~\cite{peherstorfer2018survey,patsialis2020reduced}.

\subsubsection{Relaxed solver} \label{Hp_relaxed}

The relaxed solver offers a temporal coarse-graining mechanism by solving the full nonlinear system at a coarser time resolution, typically with time steps $\Delta t' = k \Delta t$ where $k \in \mathbb{N}$ (e.g., $k = 10$ or $20$). The resulting response is linearly interpolated back to the original time grid for compatibility with the training data. This low-resolution simulation captures dominant (slow) dynamics while filtering out high-frequency nonlinear fluctuations. This approximation is particularly useful in systems where high-frequency content is not critical to performance evaluation~\cite{kramer2019nonlinear,xian2024physics}.

\subsection{Neural operator training for structural response correction} \label{Proposed_training}

With the intermediate trajectory $\vect{z}(t)$ obtained via $\mathcal{H}_p$, the neural operator $\mathcal{G}^{\mathrm{NO}}_{\vect{\theta}}$ is trained to map $\vect{z}(t)$ to the true nonlinear response $\vect{u}(t)$. The objective is to learn the residual dynamics, such as hysteresis and stiffness degradation, that are underrepresented by the simplified physical model.

The training is framed as a function-to-function regression task over the temporal domain $t \in [0, \tau]$ and across $n_d$ structural degrees of freedom. The input to the neural operator is represented as $[\vect{z}(t), t]$, combining the intermediate surrogate trajectory with the corresponding time coordinate. The architecture of the neural operator adopts the FNO formulation described in Section~\ref{NO_FNO}, comprising a lifting layer, multiple spectral convolution layers, and a final projection layer.

The training loss is defined as the relative $\mathcal{L}^2$ error across all training samples
\begin{equation} \label{Eq:Loss}
\mathcal{L}_{\mathrm{rel}}(\vect\theta) = 
\frac{ \displaystyle \sum_{j=1}^{N} \sum_{i=1}^{n_d} \left\| \hat{u}^{(j)}_i - u^{(j)}_i \right\|_2 }
{ \displaystyle \sum_{j=1}^{N} \sum_{i=1}^{n_d} \left\| u^{(j)}_i \right\|_2 } \,,
\end{equation}
where $u^{(j)}_i$ denotes the target time history response of the $i$-th degree of freedom for the $j$-th sample, and $\hat{u}^{(j)}_i$ is the predicted time history.

\section{Linear regression-based refinement of C-PhysFNO predictions} \label{Postprocessing}

To further refine the predicted structural response trajectories and provide preliminary uncertainty quantification, we introduce a postprocessing step based on linear regression. This step combines the C-PhysFNO outputs with the underlying physics-based approximation to produce a re-calibrated prediction, along with quantification of the prediction uncertainty.

For each degree of freedom $i$, we construct a linear regression model of the form
\begin{equation}
\tilde{u}_i(t) = \left[1, z_i(t), \hat{u}_i(t)\right] 
\begin{bmatrix}
  w_{i1} \\
  w_{i2} \\
  w_{i3}
\end{bmatrix}
+ \varepsilon_i(t), \quad i = 1, 2, \dots, n_d\,.
\end{equation}
Here, we use three basis functions: a constant term for potential baseline correction, $z_i(t)$ from the simplified physics-based model, and $\hat{u}_i(t)$ from the C-PhysFNO prediction. The weights $w_{i*}$ are assumed to be constants, corresponding to a stationary regression model over the time sequence. The random residual $\varepsilon_i(t)$ is assumed to be zero-mean Gaussian white noise with standard deviation $\sigma_i$. More refined residual models can be investigated in future studies. 

Model calibration is performed by randomly selecting a fixed number of response values from the existing training set of the C-PhysFNO model, and then applying the classic least-squares solution for $w_{i*}$ and $\sigma_i$. Specifically, the linear regression is trained using a dataset of the form $\{ z_i^{(l)}(t_k), \hat{u}_i^{(l)}(t_k), u_i^{(l)}(t_k) \}_{(k,l) \in \mathcal{I}}$, corresponding to randomly sampled time points $t_k$ and training inputs indexed by $l$. In this study, the size of the index set $\mathcal{I}$ is set to 1000.

Once trained, the regression model yields the refined prediction and its associated uncertainty via the estimated mean $\mathbb{E}[\tilde{u}_i(t)] = w_{i1} + w_{i2} z_i(t) + w_{i3} \hat{u}_i(t)$ and variance $\mathrm{Var}[\tilde{u}_i(t)] = \sigma_i^2$. It is worth restating that this linear regression is purely postprocessing, as both the training and prediction phases do not incur additional calls to the original model, the simplified model, or the C-PhysFNO model.

\section{Numerical investigations} \label{Examples}

This section evaluates the proposed C-PhysFNO framework and its linear regression-based refinement strategy using three representative structural examples: (1) a five-story building with Bouc–Wen hysteresis subjected to white noise excitation, (2) a prestressed concrete bridge model under synthetic ground motions, and (3) a nine-story steel moment-resisting frame subjected to recorded earthquake inputs.

To quantify prediction accuracy, two error metrics are defined for each degree of freedom $i$: the root-mean-square error (RMSE) and the relative $\mathcal{L}^2$ error. These are computed as
\begin{equation} \label{Eq:RMSE}
\varepsilon_{i}^{\mathrm{RMSE}} = \sqrt{\frac{1}{N_t n_t}\sum_{j=1}^{N_t}\sum_{k=1}^{n_t}\left( \hat{u}_i^{(j)}(t_k)-u_i^{(j)}(t_k) \right)^2} ,
\end{equation}
\begin{equation} \label{Eq:Relative}
\varepsilon_{i}^{\mathrm{Rel}} = \sqrt{ \frac{\displaystyle \sum_{j=1}^{N_t}\sum_{k=1}^{n_t}\left( \hat{u}_i^{(j)}(t_k)-u_i^{(j)}(t_k) \right)^2} {\displaystyle \sum_{j=1}^{N_t}\sum_{k=1}^{n_t}\left( u_i^{(j)}(t_k) \right)^2} } ,
\end{equation}
where $N_t$ is the number of testing samples and $n_t$ denotes the number of temporal discretization points used to represent each response trajectory. These metrics assess both absolute and scale-invariant errors, providing a comprehensive basis for comparing different  approaches.

\subsection{Example 1: Nonlinear system with Bouc–Wen hysteresis} \label{MDOF_BW}

\subsubsection{Structural model and stochastic excitation}

\begin{figure}[H]
  \centering
  \includegraphics[scale=0.50] {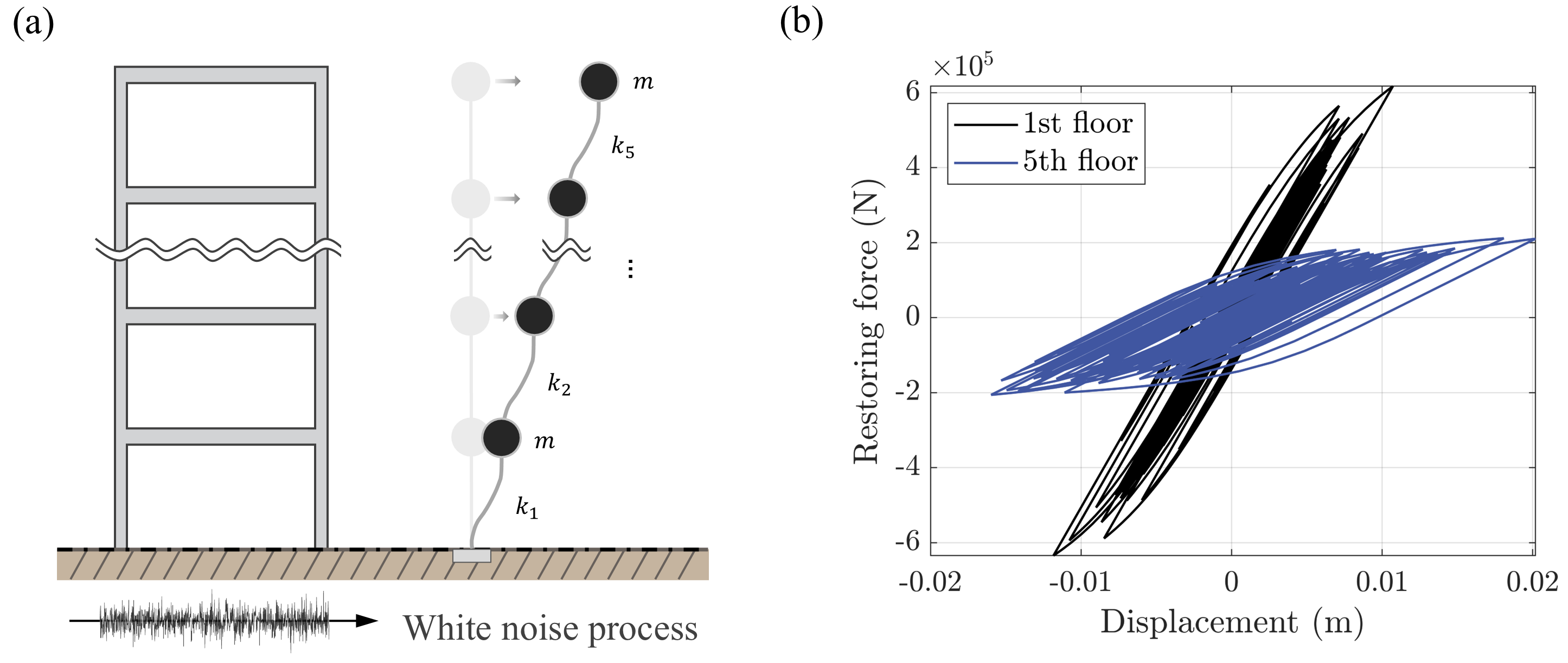}
  \caption{\textbf{(a) Five-story nonlinear building system and (b) hysteresis loops at the 1st and 5th stories.}}
  \label{Fig_MDOF_BW}
\end{figure}

The first example considers a five-story nonlinear shear building model, shown in Figure~\ref{Fig_MDOF_BW}, with restoring forces governed by the Bouc–Wen hysteresis model~\cite{song2006generalized,ismail2009hysteresis,kim2024dimensionality}. The hysteretic force for the $i$-th story is expressed as:
\begin{equation} \label{Eq:BW_eq1}
r_i(t) = k_i[\alpha_i v_i(t) + (1-\alpha_i)h_i(t)] \,,
\end{equation}
\begin{equation} \label{Eq:BW_eq2}
\dot{h}_i(t) =  -\delta|\dot{v}_i(t)||h_i(t)|^{(\bar{n}-1)}h_i(t) - \zeta\dot{v}_i(t)|h_i(t)|^{\bar{n}} + A\dot{v}_i(t)\,,  \,\,\,  i=1,...,5 \,, 
\end{equation}
where $v_i(t)$ is the local interstory deformation, $r_i(t)$ is the shear restoring force, $k_i$ is the elastic stiffness, and $h_i(t)$ is the hysteretic response governed by Eq.~\eqref{Eq:BW_eq2}. The deformation vector $\vect{v}(t)$ is obtained from the global displacement vector $\vect{u}(t)$ via a compatibility matrix $\vect{A}_{\vect{f}}$, i.e., $\vect{v}(t) = \vect{A}_{\vect{f}} \vect{u}(t)$. The Bouc–Wen parameters are set as $\alpha_i = 0.1$, $\bar{n} = 3$, $A = 1$, and $\delta = \zeta = 1/(2 u_y^{\bar{n}})$, where $u_y = 0.01$ m is the yield displacement. Each floor has a mass of $m = 3.0 \times 10^4$ kg, and damping is applied to achieve 5\% critical damping in all modes.

The base excitation $a_g(t)$ is modeled as a white noise process synthesized in the frequency domain:
\begin{equation} \label{WN_eq1}
a_{g}(t) = \sigma\sum_{j=1}^{n/2}\left[X_j\cos{(\omega_{j}t)} + X_{(n/2+j)}\sin{(\omega_{j}t)} \right] \,,
\end{equation}
where $X_j$ are independent standard Gaussian variables, $\omega_j = j \Delta \omega$, and $\sigma = \sqrt{2 S \Delta \omega}$ with $S = 0.015$ m$^2$/s$^3$ representing the spectral intensity. The discretization parameters are set as $n = 1200$, $\Delta \omega = \pi/30$, and the cutoff frequency $\omega_{\text{cut}} = 20\pi$ rad/s. The resulting excitation signal is sampled at $\Delta t = 0.01$ s over a 30-second duration, yielding 3,001 time steps.

\subsubsection{Prediction performance and discussion} \label{MDOF_BW:results}

We compare the prediction performance of the standard FNO model with the proposed C-PhysFNO framework using three distinct coarse-graining strategies for constructing the intermediate representation $\mathcal{H}_p$: (i) ELS, (ii) modal reduction, and (iii) relaxed solver. The details of each $\mathcal{H}_p$ are as follows:
\begin{itemize}
    \item ELS: The response is obtained by solving Eq.\eqref{Eq:ELS} with the effective stiffness matrix $\mathbf{K}_{\mathrm{eq}}$ defined using initial elastic stiffness values $k_i$ in Eq.\eqref{Eq:BW_eq1}. The mass and damping matrices are retained from the original nonlinear system.
    \item Modal reduction: The system is reduced using the first $r = 2$ mode shapes. The Bouc–Wen restoring force is projected onto this reduced modal basis using the formulations provided in Eqs.~\eqref{Eq:modal_projected}–\eqref{Eq:modal_projected2}.
    \item Relaxed solver: The full nonlinear system is solved using a coarser temporal discretization with time step $\Delta t' = 30\Delta t$, generating trajectories of length 101. These are linearly interpolated onto the original fine time grid to match the resolution of the target trajectories.
\end{itemize}

All models are trained using 400 training samples, validated on 100 samples, and evaluated on 200 test samples. Both standard FNO and C-PhysFNO use 6 Fourier kernel layers, 32 Fourier modes, and 64 feature channels. In the standard FNO, the lifting and projection operators are implemented as fully connected layers that transform the input feature channels via the mappings $2 \rightarrow 64$ and $64 \rightarrow 128 \rightarrow n_d$ (with $n_d=5$ in this example), respectively, where the input dimension $2=1+1$ corresponds to the ground acceleration $a_g(t)$ and the time coordinate $t$. For C-PhysFNO, the input channel dimension starts with $n_d+1=6$, including the intermediate representation and time, and the network follows the mapping $6 \rightarrow 64$ and $64 \rightarrow 128 \rightarrow n_d$. Training is performed using the Adam optimizer with an initial learning rate of 0.001 and a decay factor of 0.5. Training of the C-PhysFNO model in this example took approximately 12 minutes using an NVIDIA GeForce RTX 4070 Laptop GPU.

\begin{figure}[H]
  \centering
  \includegraphics[scale=0.77] {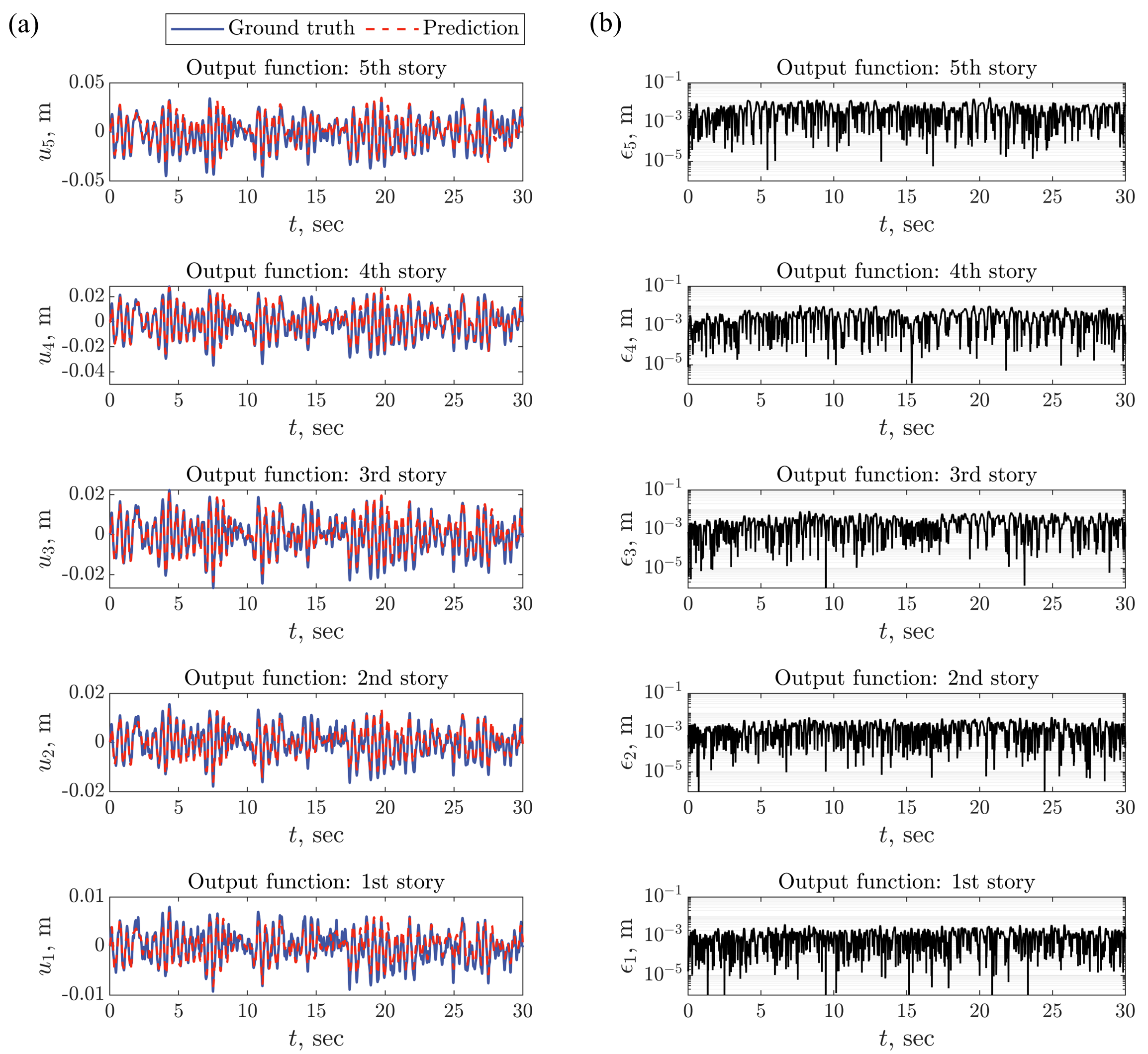}
  \caption{\textbf{Predicted structural responses using C-PhysFNO with ELS for the five-story nonlinear system under white noise excitation: (a) displacement trajectories for all stories, and (b) corresponding  absolute prediction error.}}
  \label{Fig_pred_BW}
\end{figure}

\begin{figure}[H]
  \centering
  \includegraphics[scale=0.50] {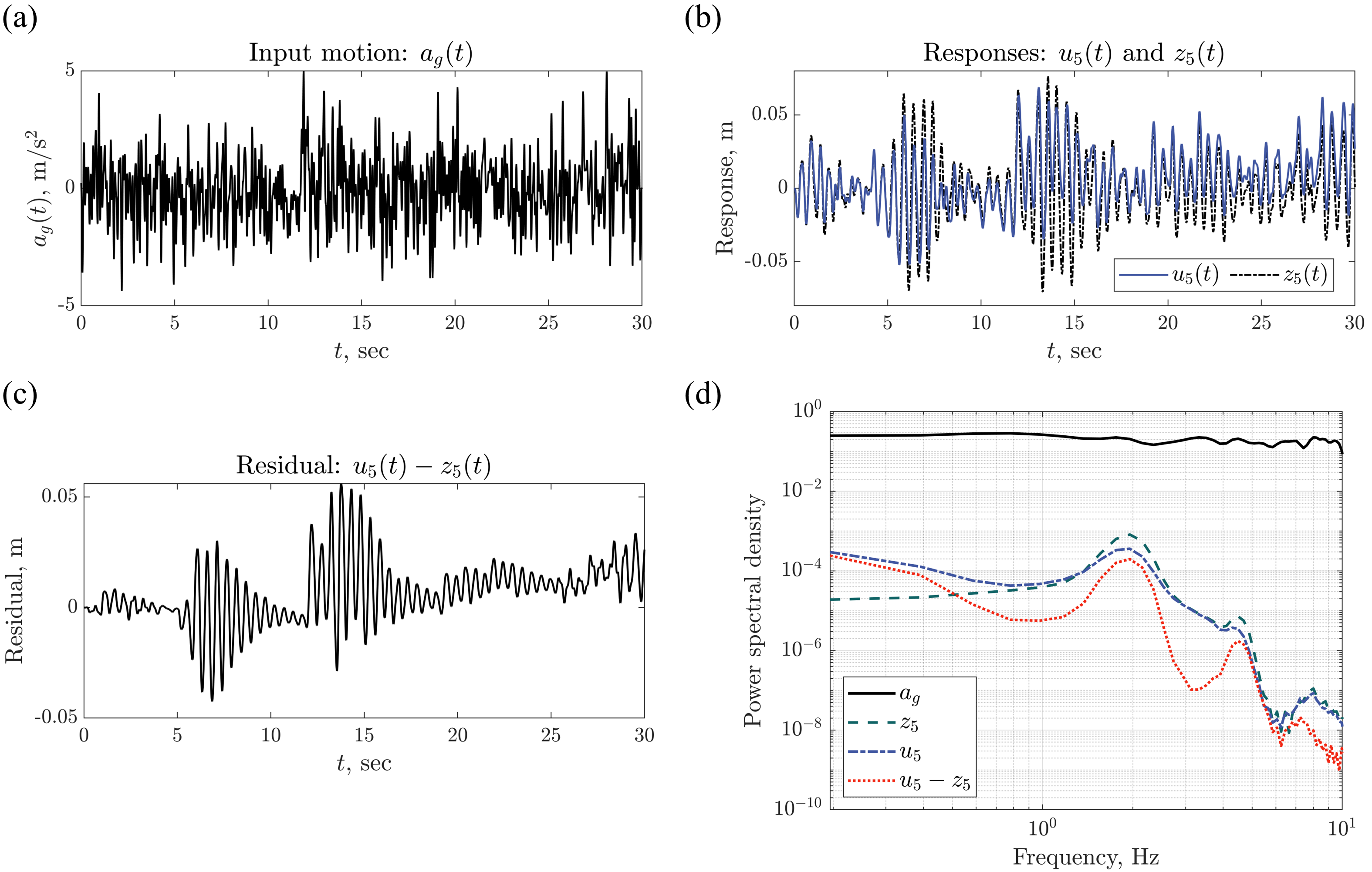}
  \caption{\textbf{Time-domain and frequency-domain analysis of representative top-floor response: (a) Ground acceleration $a_g(t)$, (b) intermediate response $z(t)$ and true nonlinear response $u(t)$, (c) residual $r(t) = u(t) - z(t)$, and (d) PSDs of all signals.}}
  \label{Fig_BW_Residual_PSD}
\end{figure}

Figure~\ref{Fig_pred_BW} presents an example of the predicted displacement trajectories and corresponding absolute errors, i.e., $\epsilon_i=|u_i-\hat{u}_i|$, for a representative test case using C-PhysFNO with the ELS as $\mathcal{H}_p$. The results exhibit close agreement with the ground truth nonlinear responses obtained from finite element simulations with errors consistently below 0.01 m. This demonstrates the capability of the proposed model to accurately reconstruct trajectory-level dynamics under wide-band excitations. To further assess how the intermediate representation mitigates spectral bias \cite{rahaman2019spectral}, we examine the frequency-domain characteristics of the relevant signals. Figure~\ref{Fig_BW_Residual_PSD} presents the time histories and power spectral densities (PSDs) for: (i) the input ground acceleration $a_g(t)$, (ii) the intermediate response $z(t)$ from the ELS model and the true nonlinear response $u(t)$, and (iii) the residual $r(t)=u(t)-z(t)$, for a representative top-story response. The PSDs are estimated using Welch’s method. While the residual retains some high-frequency components, its spectral content is substantially reduced and more localized compared to the original nonlinear response. This suggests that the residual mapping learned by the FNO is spectrally simpler and statistically better conditioned. By learning to correct the residual rather than the full mapping, the C-PhysFNO reduces the spectral gap between input and output, aligns the encoded inputs with the dominant solution modes, and improves learning efficiency. These results support the hypothesis that physics-informed intermediate representations mitigate spectral bias and enhance the data efficiency and generalization capability of neural operators. It is worth noting that PSD captures second-order statistics of the signal and cannot fully describe nonlinear effects such as hysteresis or rare extreme responses. Therefore, similarity in PSDs between ELS and nonlinear responses does not imply equivalence in trajectory-level dynamics.

\begin{figure}[H]
  \centering
  \includegraphics[scale=0.55] {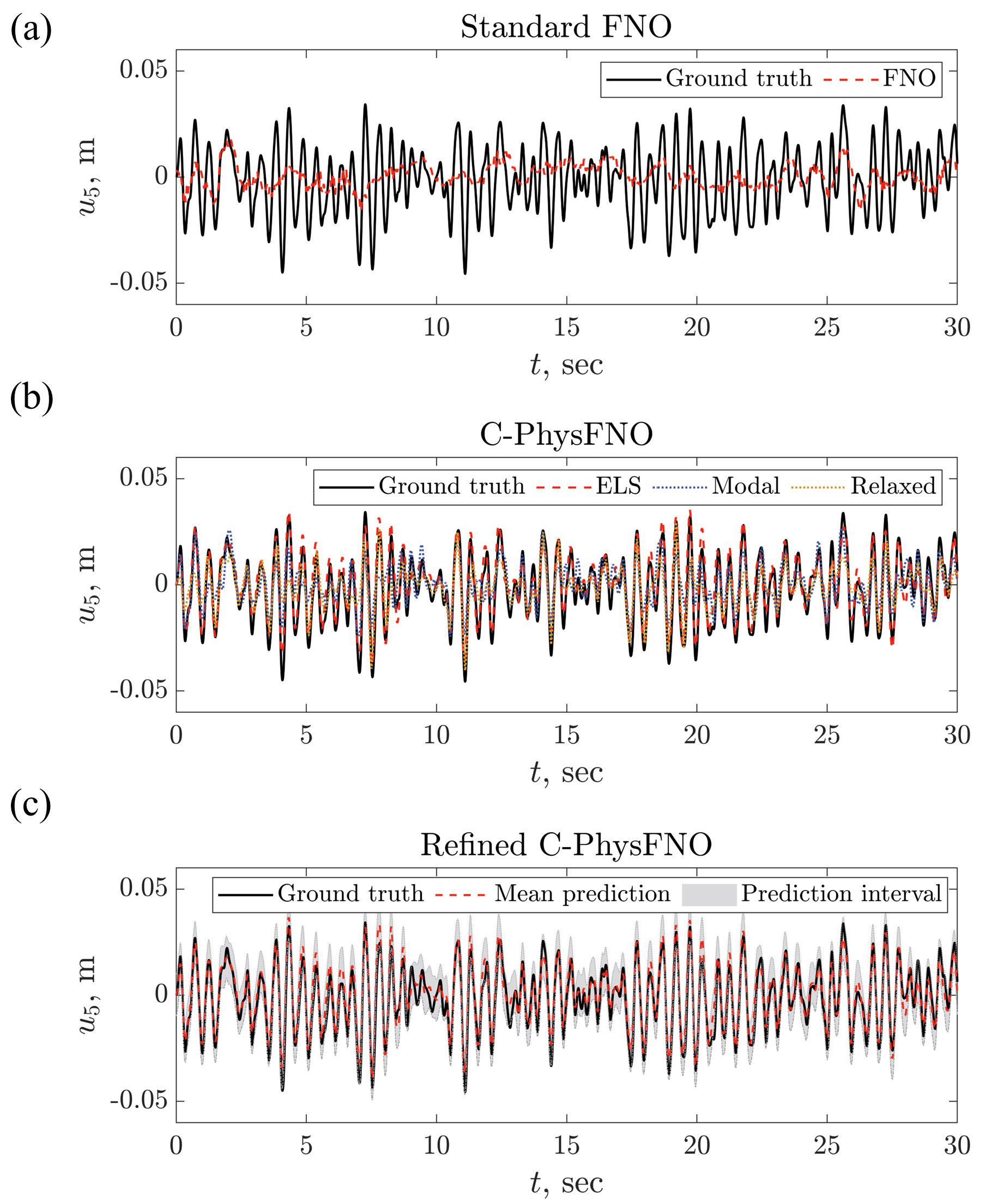}
  \caption{\textbf{Comparison of top-story response predictions for the five-story nonlinear system: (a) predictions using standard FNO, (b) predictions using C-PhysFNO with three coarse-graining strategies, and (c) predictions from linear regression-refined C-PhysFNO (ELS case).} All models are evaluated under the same ground motion input. In (c), the shaded area represents the prediction interval given by the posterior mean $\pm$ one standard deviation.}
  \label{Fig_compare_BW}
\end{figure}

Figure~\ref{Fig_compare_BW} compares top-story response predictions across all methods under the same test input: the standard FNO, C-PhysFNO with three different coarse-graining strategies, and the linear regression-based refinement applied to the ELS case. The standard FNO fails to capture key response trends, whereas C-PhysFNO accurately reconstructs trajectory-level dynamics across all three $\mathcal{H}_p$ types. These observations support the hypothesis that broadband stochastic excitations (such as band-limited white noise) lead to highly variable and spectrally complex input–output mappings, which are difficult to learn directly. By introducing a physics-based intermediate representation, C-PhysFNO effectively reduces this spectral gap, simplifying the learning task into a better-conditioned residual regression problem. Postprocessing via linear regression further refines the predictions and quantifies confidence intervals. Notably, the white noise excitation poses a challenging prediction task due to the lack of structure, highlighting the benefit of injecting an intermediate physics-based model.

Table~\ref{Tab_result_BW} presents the RMSE and relative $\mathcal{L}^2$ error for all discussed neural operator models, with metrics averaged over 10 independent runs with different random seed. For the refined models, error metrics by mean predictions are reported. The results indicate that all C-PhysFNO variants consistently outperform the standard FNO, with the ELS-based model achieving the highest accuracy among the inspected coarse-graining strategies. Postprocessing via linear regression further refines the predictions, yielding modest yet consistent improvements across all C-PhysFNO variants.

\begin{figure}[H]
  \centering
  \includegraphics[scale=0.50] {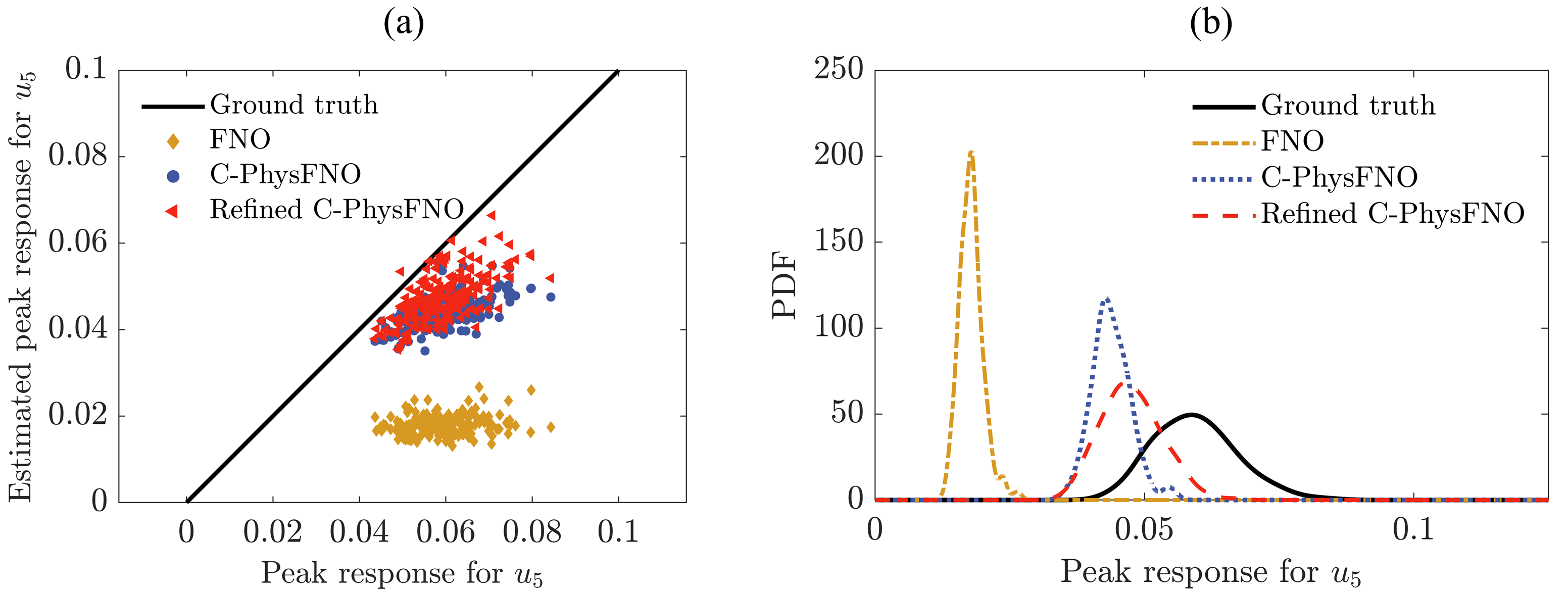}
  \caption{\textbf{Top-story peak displacement prediction for the five-story nonlinear system: (a) scatter plot of predicted versus ground truth peak values across the test set, and (b) estimated probability distribution of peak responses.}}
  \label{Fig_BW_Peak}
\end{figure}

Figure~\ref{Fig_BW_Peak} presents a scatter plot and probability distribution of the top-story peak displacement during the response time window, evaluated across the test set. C-PhysFNO and its refinement more accurately match the ground truth distribution compared to standard FNO. Nonetheless, notable discrepancies remain in capturing extreme values. This limitation is partly attributed to the nature of the model, which is trained to predict time-domain response trajectories rather than directly estimating peak values. In this example, the broadband excitation introduces high-frequency components and non-stationary behavior, making peak response estimation particularly challenging. Moreover, small misalignments in predicted trajectories can lead to amplified discrepancies in peak statistics. These observations suggest that future research incorporating Bayesian neural operators~\cite{bulte2025probabilistic} with active learning could offer improved tail behavior modeling and rare events.

\begin{table}[H]
  \caption{\textbf{Prediction error metrics for the five-story nonlinear system across all neural operator models.}}
  \label{Tab_result_BW}
  \centering
  \resizebox{\textwidth}{!}{  
  \begin{tabular}{l l | c c c c c | c c c c c}
    \toprule
    \multirow{2}{*}{Model} & \multirow{2}{*}{\shortstack{Intermediate\\representation}} & \multicolumn{5}{c|}{RMSE (m)} & \multicolumn{5}{c}{Relative $\mathcal{L}^2$ error} \\ \cmidrule(lr){3-12}
     & & Story 1 & Story 2 & Story 3 & Story 4 & Story 5 & Story 1 & Story 2 & Story 3 & Story 4 & Story 5 \\   
    \midrule
    FNO & - & 0.0038  &  0.0072  &  0.0105 &   0.0140  &  0.0187 & 0.9653  &  0.9569  &  0.9561  &  0.9578  &  0.9582 \\
    \midrule
    \multirow{3}{*}{C-PhysFNO} & ELS & 0.0022  &  0.0040   & 0.0054  &  0.0070 & 0.0102 & 0.5775 & 0.5314  &  0.4957  &  0.4819  &  0.5245 \\
     & Model reduction & 0.0028 &   0.0053 &   0.0077  &  0.0103 &   0.0139 & 0.7196 &   0.7059 &   0.7025  &  0.7034 &   0.7134 \\
     & Relaxed solver & 0.0025  &  0.0048  &  0.0070  &  0.0094  &  0.0126 & 0.6398  &  0.6393 &   0.6404 &   0.6421 &   0.6485 \\
    \midrule
    \multirow{3}{*}{\shortstack{Refined\\C-PhysFNO}} & ELS & 0.0021  &  0.0039   & 0.0053  &  0.0069 & 0.0100 & 0.5492 & 0.5200  &  0.4882  &  0.4780  &  0.5208 \\
     & Model reduction & 0.0027  &  0.0052   & 0.0076  &  0.0102 & 0.0138 & 0.7015 & 0.6820  &  0.6832  &  0.6850  &  0.7002 \\ 
     & Relaxed solver & 0.0024  &  0.0047   & 0.0069  &  0.0093 & 0.0125 & 0.6180 & 0.6282  &  0.6302  &  0.6310  &  0.6402 \\  
    \bottomrule
  \end{tabular} }  \\
\end{table}

\subsubsection{Parametric study}

\begin{figure}[H]
  \centering
  \includegraphics[scale=0.46] {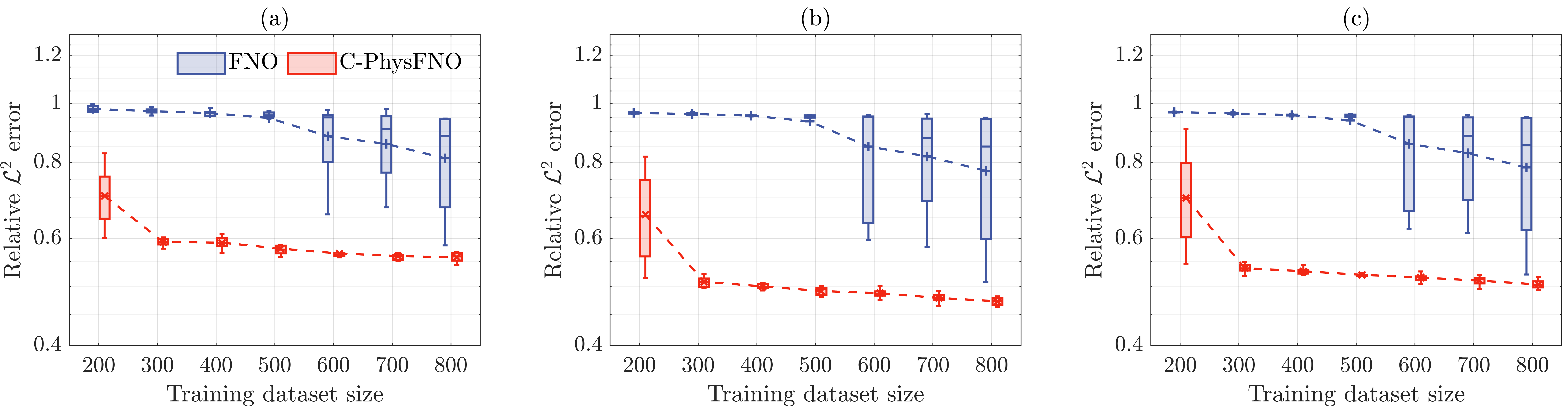}
  \caption{\textbf{Relative $\mathcal{L}^2$ error across varying training dataset sizes: (a) 1st, (b) 3rd, and (c) 5th story predictions.} Each box represents results from 10 independent runs.}
  \label{Fig_error_NT}
\end{figure}

\begin{figure}[H]
  \centering
  \includegraphics[scale=0.46] {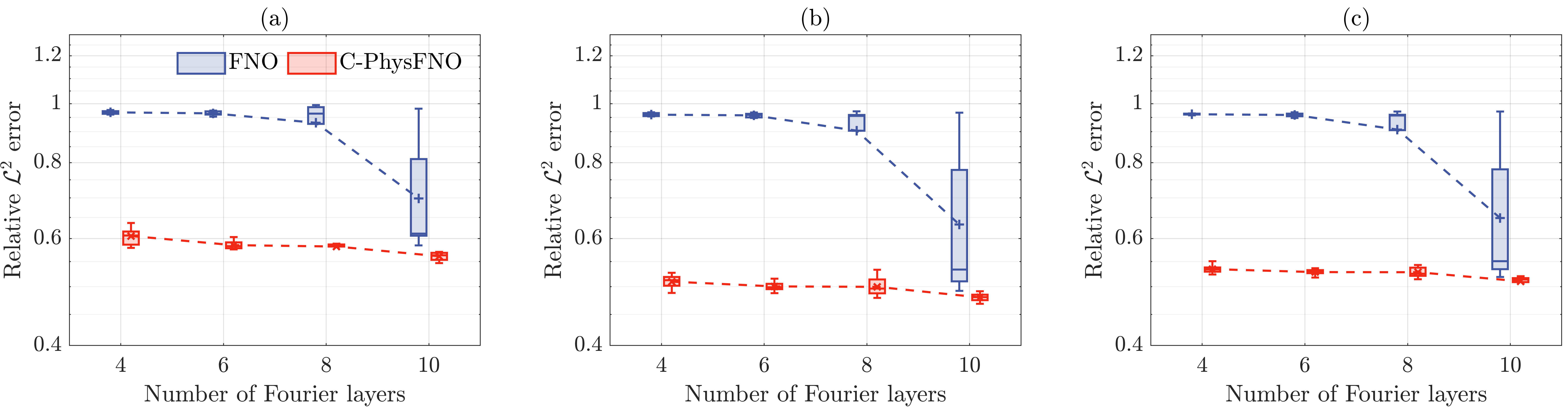}
  \caption{\textbf{Relative $\mathcal{L}^2$ error across different number of Fourier layers: (a) 1st, (b) 3rd, and (c) 5th story predictions.} Each box represents results from 10 independent runs.}
  \label{Fig_error_FL}
\end{figure}

A parametric study is conducted to evaluate neural operator model sensitivity to training dataset size and the number of Fourier kernel layers. Figures~\ref{Fig_error_NT} and \ref{Fig_error_FL} show relative $\mathcal{L}^2$ errors across 10 runs for varying configurations. These experiments are conducted for C-PhysFNO using ELS as $\mathcal{H}_p$, without postprocessing. While the performance of the standard FNO improves with increased data availability and deeper architectures, the C-PhysFNO consistently achieves lower error across all cases. In fact, the standard FNO achieves competitive accuracy when sufficient training data or model depth is provided, especially in smoother or more structured input–output settings. Notably, the performance gap is most pronounced under data-limited conditions, highlighting the advantage of integrating domain knowledge through physics-based intermediate representations.


\subsection{Example 2: Auburn Ravine bridge subjected to synthetic ground motions} \label{Ex_ARB}

\subsubsection{Structural model and ground motion dataset}

This example investigates the seismic response of the Auburn Ravine bridge, a prestressed concrete box-girder structure comprising six spans and five bents, each supported by two circular piers~\cite{konakli2011stochastic}. A detailed finite element model is developed in OpenSees~\cite{mckenna2011opensees}, incorporating geometric nonlinearity and material nonlinearities in the pier columns and connection regions. The total bridge length is 166.4 m, and key structural properties are summarized in Table~\ref{Tab_ARbridge}. The piers are modeled using force-based beam-column elements with fiber-section discretization. Confined and unconfined concrete fibers and reinforcing steel bars represent axial–bending interaction, while shear and torsion are modeled using uniaxial components. The deck is rigidly connected to the piers, which are fixed at their bases. Translational springs are used to model abutment stiffness. The fundamental period of the bridge is $T = 0.38$ s. The quantity of interest is the displacement time history at the top of each of the ten piers, denoted by $\vect{u}(t) = \{u_1(t), \dots, u_{10}(t)\}$.

\begin{figure}[H]
  \centering
  \includegraphics[scale=0.48] {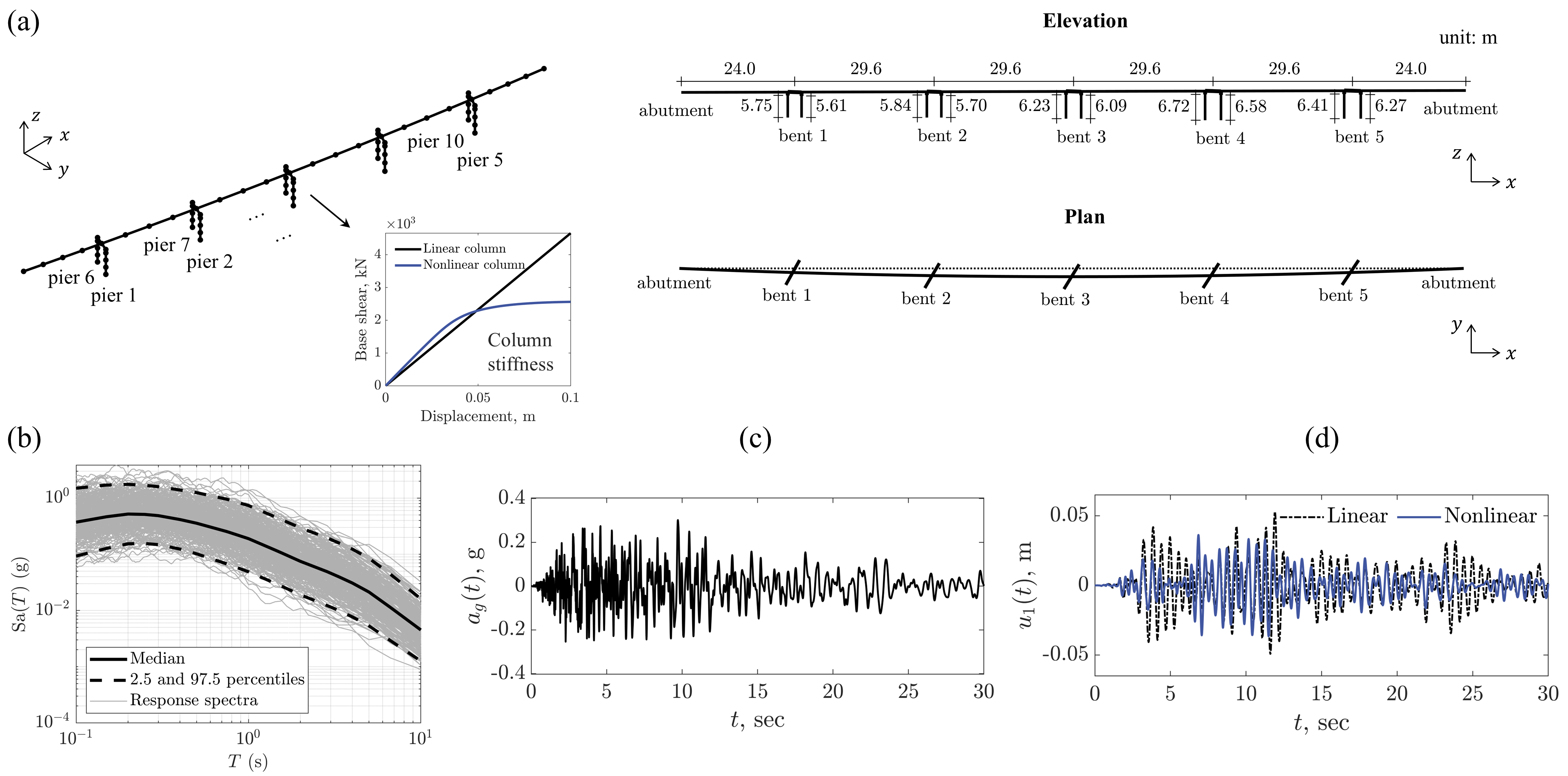}
  \caption{\textbf{(a) Structural model of the Auburn Ravine bridge, (b) response spectra of the synthetic ground motions, (c) a representative input acceleration time history, and (d) top displacement response at Pier 1.} In (b), the median and 2.5\%-97.5\% quantiles of the target spectrum are superimposed. In (d), both the full nonlinear response and the corresponding physics-based approximation using the linearized model are shown.}
  \label{Fig_ARB}
\end{figure}

The ground motion dataset consists of spectrum-compatible synthetic accelerograms generated using the frequency-domain spectrum-matching algorithm~\cite{yanni2024probabilistic}. The target response spectrum is defined using the Next Generation Attenuation (NGA)-West2 ground motion model~\cite{boore2014nga} and the empirical spectral correlation structure~\cite{baker2008correlation}. Each synthetic accelerogram is constructed using a modulated Fourier series with iterative amplitude adjustment to match the target spectrum, followed by baseline correction to remove long-period drift. A total of 1000 synthetic ground motions are generated with a time step of $\Delta t = 0.01$ s and a duration of 30 s, resulting in 3,001 time steps per sample. Figure~\ref{Fig_ARB} illustrates the finite element model, the response spectra of the synthetic motions, and representative structural responses.

\begin{table}[H]
  \caption{\textbf{Structural parameters for the Auburn Ravine bridge model.}}
  \label{Tab_ARbridge}
  \centering
  \begin{tabular}{c c}
    \toprule
    Parameter & Value \\
    \midrule
    Damping ratio, \% & 5 \\
    Cross-sectional area (girder), m$^2$ & 6.43 \\
    Elastic modulus (girder), GPa & 28.3 \\
    Elastic modulus reinforcing steel (pier), GPa & 200 \\
    Yield strength of reinforcing steel (pier), MPa & 475 \\
    Ultimate strength of reinforcing steel (pier), MPa & 655 \\
    Onset strain of steel hardening (pier) & 0.0115 \\
    Elastic modulus of concrete (pier), GPa & 27.6 \\
    Compressive strength of concrete (pier), MPa & 34.5 \\
    Strain at compressive strength of concrete (pier) & 0.002 \\
    Column diameter, m & 1.38 \\
    Concrete cover thickness, m & 0.05 \\
    \bottomrule
  \end{tabular}  \\
\end{table}

\subsubsection{Prediction performance and discussion}

To construct the physics-based intermediate representation $\mathcal{H}_p$, a linearized version of the bridge model is developed by replacing the nonlinear fiber-section columns with elastic beam-column elements. The elastic stiffness of each pier is computed based on pre-yield sectional properties to match the initial stiffness of the nonlinear model. This linear surrogate retains the original mass, geometry, and boundary conditions but omits nonlinear behaviors such as hysteresis, stiffness degradation, and strength loss. An example trajectory from the linearized model is shown in Figure~\ref{Fig_ARB}(d). This representation provides a computationally efficient approximation of global elastic dynamics and serves as the physics-based input to the C-PhysFNO framework.

The training, validation, and testing datasets consist of 600, 100, and 300 synthetic ground motion samples, respectively, with model training completed in approximately 40 minutes. The neural operator architecture used in this example comprises 8 Fourier layers, 40 retained Fourier modes, and 128 feature channels. The lifting and projection layers follow the configuration used in Section~\ref{MDOF_BW:results}.

\begin{figure}[H]
  \centering
  \includegraphics[scale=0.95] {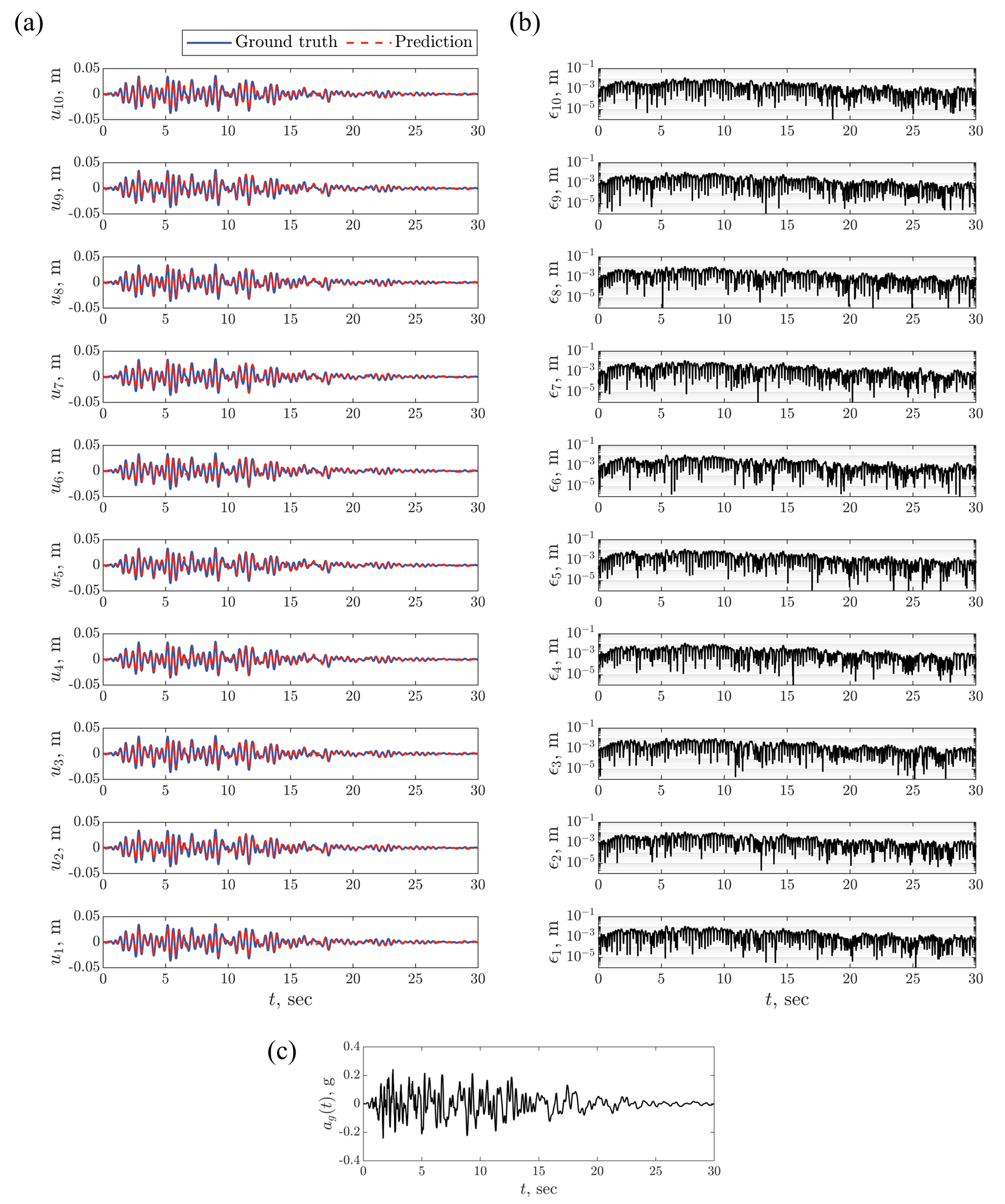}
  \caption{\textbf{Structural response predictions for the Auburn Ravine bridge using C-PhysFNO: (a) displacement trajectories at the top of each pier, (b) absolute prediction error, and (c) input ground motion.}}
  \label{Fig_pred_ARB}
\end{figure}

\begin{figure}[H]
  \centering
  \includegraphics[scale=0.52] {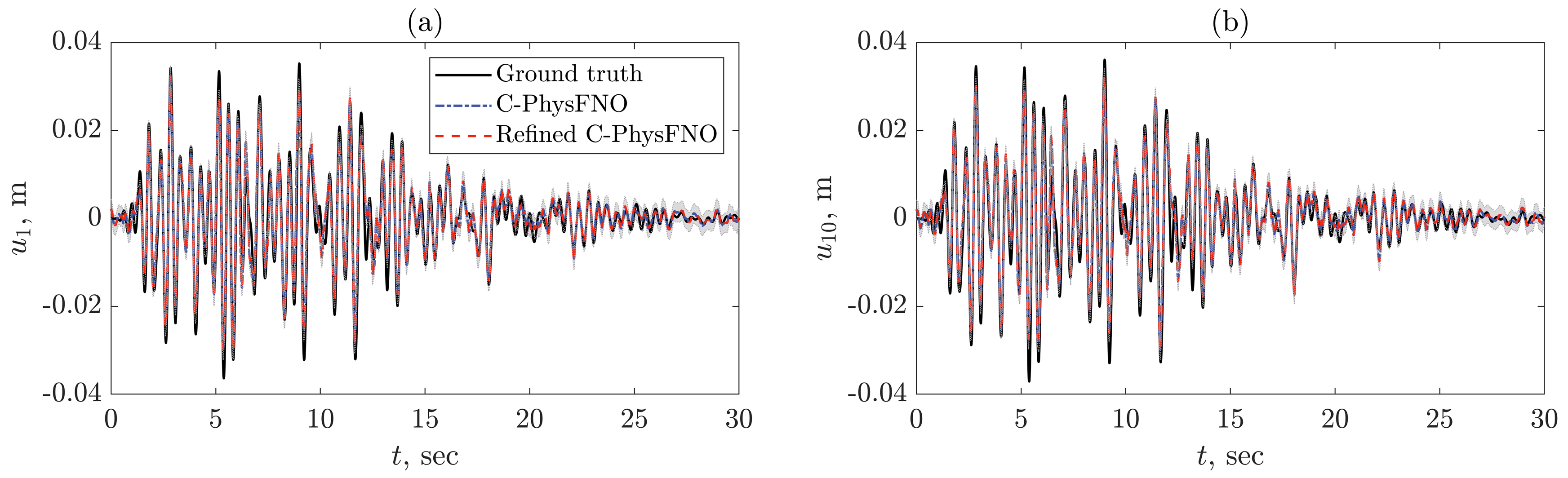}
  \caption{\textbf{Linear regression-based refinement of C-PhysFNO predictions for the Auburn Ravine bridge: (a) Pier 1, (b) Pier 10.} The shaded region represents the prediction interval defined by the posterior mean $\pm$ one standard deviation.}
  \label{Fig_Post_ARB}
\end{figure}

Figure~\ref{Fig_pred_ARB} illustrates the predicted displacement trajectories and absolute errors for all piers in a representative test case. The C-PhysFNO model accurately reconstructs nonlinear dynamics, achieving prediction errors typically below 0.001 m. Figure~\ref{Fig_Post_ARB} shows the linear regression-based refinement for Pier 1 and Pier 10. The refined predictions offer smoother trajectories and quantify uncertainty through posterior intervals, although their accuracy remains comparable to the original C-PhysFNO predictions.

Table~\ref{Tab_result_ARB} summarizes the RMSE and relative $\mathcal{L}^2$ error across all ten piers, averaged over 10 independent training runs. C-PhysFNO and its linear regression-based refinement consistently outperform the standard FNO baseline. However, it is worth noting that the standard FNO also achieves reasonably accurate predictions in this example, due in part to the spectral compatibility and smoother structure of the input–output mappings. The refinement does not improve the C-PhysFNO prediction, indicating that its primary contribution in this example lies in a preliminary uncertainty quantification for the prediction.

\begin{figure}[H]
  \centering
  \includegraphics[scale=0.50] {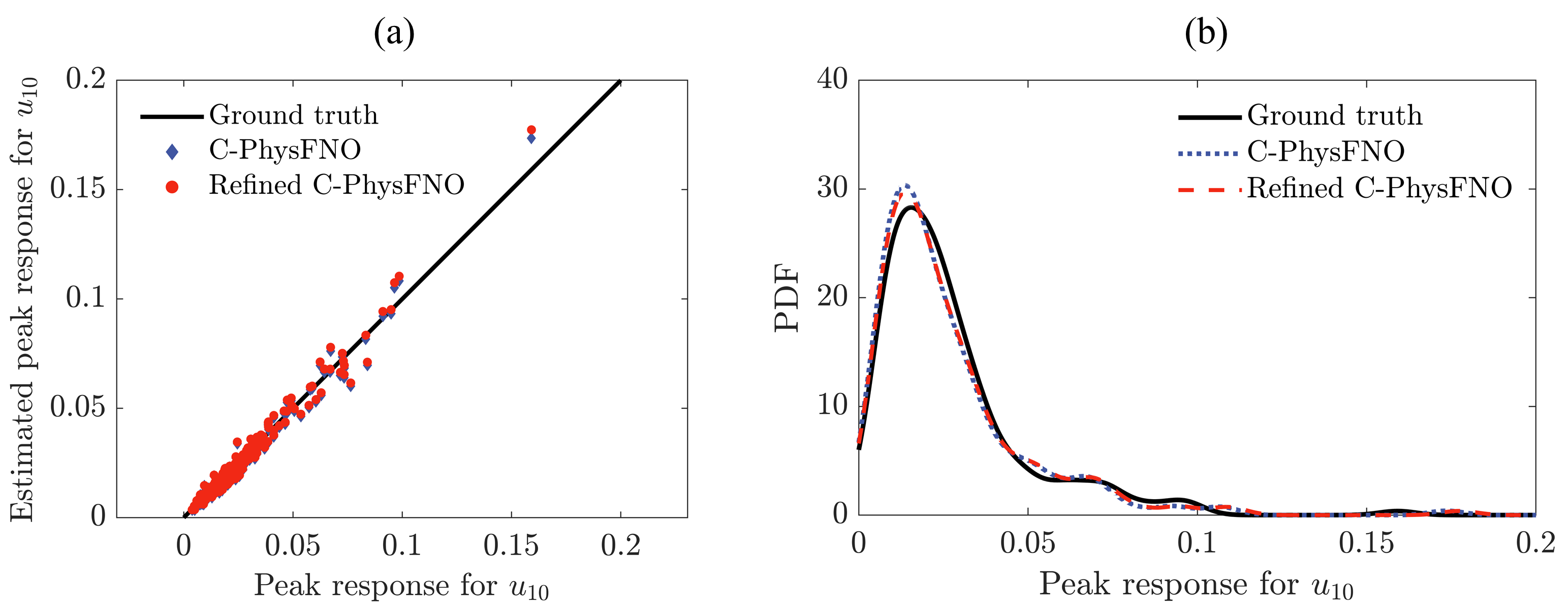}
  \caption{\textbf{Top-story peak displacement prediction for the Auburn Ravine bridge: (a) scatter plot of predicted versus ground truth peak values across the test set, and (b) estimated probability distribution of peak responses.} Both C-PhysFNO and its refinement provide accurate predictions with $R^2$ values of 0.9730 and 0.9721, respectively.}
  \label{Fig_ARB_Peak}
\end{figure}

In addition, Figure~\ref{Fig_ARB_Peak} demonstrates that both C-PhysFNO and its refinement produce accurate estimates of peak displacements, effectively capturing not only the central tendency but also the overall distribution of the peak response. This performance is likely supported by the use of spectrally compatible input motions, which mitigate frequency mismatch and thus alleviate challenges in the operator learning.

\begin{table}[H]
  \caption{\textbf{Prediction error metrics for the Auburn Ravine bridge across all neural operator models.}}
  \label{Tab_result_ARB}
  \centering
  \resizebox{\textwidth}{!}{
  \begin{tabular}{l | l | c c c c c c c c c c}
    \toprule
    Model & Error metric & Pier 1 & Pier 2 & Pier 3 & Pier 4 & Pier 5 & Pier 6 & Pier 7 & Pier 8 & Pier 9 & Pier 10 \\
    \midrule
    \multirow{2}{*}{FNO} 
    & RMSE (m) & 0.0061  &  0.0060 &   0.0059  &  0.0059  &  0.0057  &  0.0059  &  0.0060  &  0.0061 &   0.0062  &  0.0062 \\
    & Relative $\mathcal{L}^2$ & 0.6402  &  0.6415 &  0.6370 &   0.6385  &  0.6343 &  0.6436 &   0.6405  &  0.6439 &  0.6398 &    0.6416 \\
    \midrule
    \multirow{2}{*}{C-PhysFNO} 
    & RMSE (m) & 0.0046 & 0.0047 & 0.0045 & 0.0045 & 0.0044 & 0.0046 & 0.0046 & 0.0046 & 0.0047 & 0.0047 \\
    & Relative $\mathcal{L}^2$ & 0.4873 & 0.4939 & 0.4845 & 0.4919 & 0.4855 & 0.4994 & 0.4930 & 0.4895 & 0.4882 & 0.4903 \\
    \midrule
    \multirow{2}{*}{\shortstack{Refined\\C-PhysFNO}} 
    & RMSE (m) & 0.0046 & 0.0047 & 0.0046 & 0.0046 & 0.0045 & 0.0047 & 0.0047 & 0.0047 & 0.0047 & 0.0048 \\
    & Relative $\mathcal{L}^2$ & 0.4902 & 0.4980 & 0.4880 & 0.4939 & 0.4900 & 0.5045 & 0.4992 & 0.4945 & 0.4912 & 0.4955 \\    
    \bottomrule
  \end{tabular}}
\end{table}

\subsection{Example 3: Nine-story steel moment-resisting frame under recorded ground motions} \label{Ex_SAC9}

\subsubsection{Structural model and ground motion dataset}

\begin{figure}[H]
  \centering
  \includegraphics[scale=0.48] {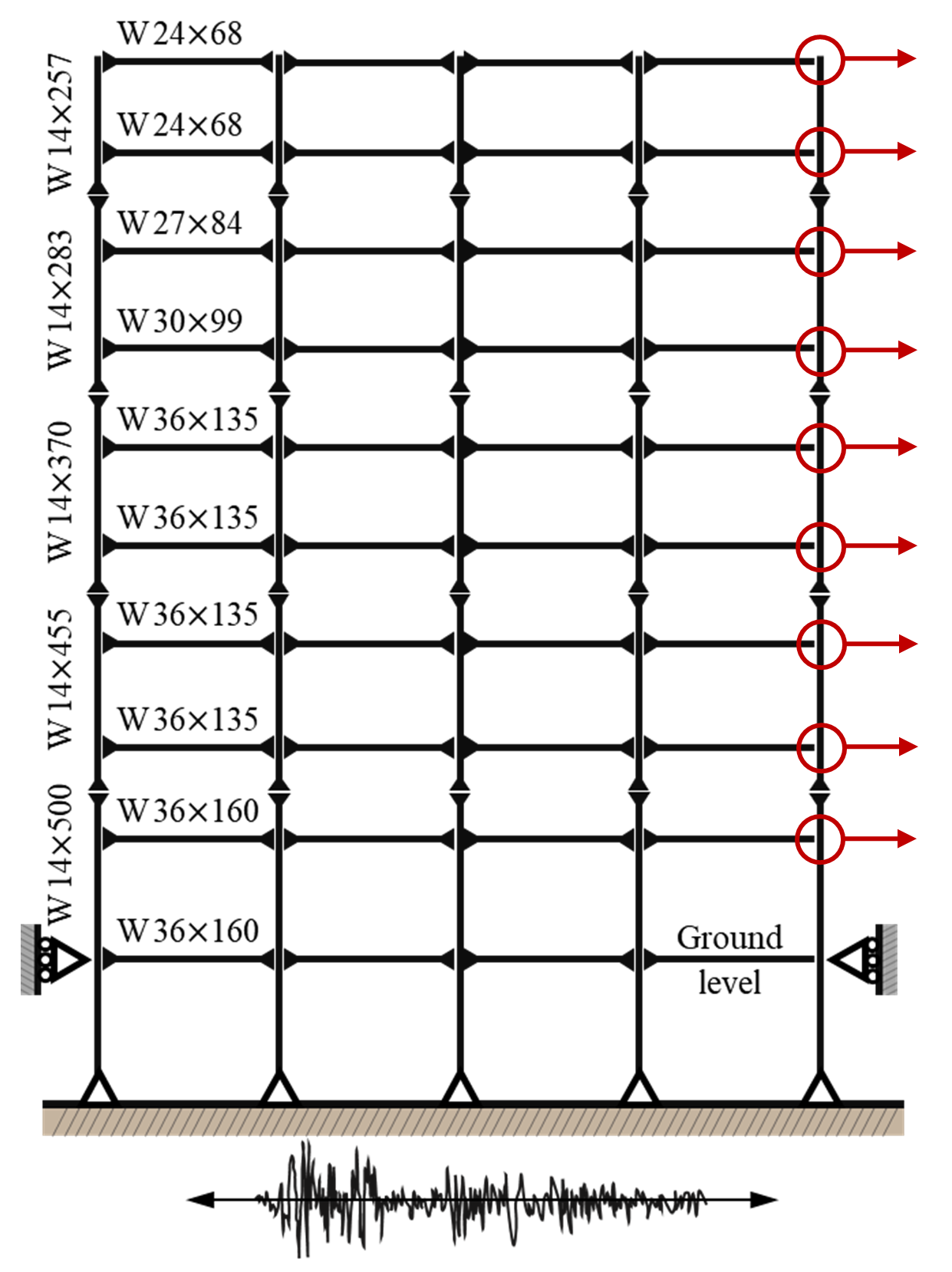}
  \caption{\textbf{Nine-story steel moment-resisting frame.} The red circles indicate the nodes where displacement responses $\vect{u}(t)$ are extracted.}
  \label{Fig_SAC9}
\end{figure}

This example evaluates the proposed framework using a nine-story steel moment-resisting frame from the SAC Joint Venture benchmark study~\cite{ohtori2004benchmark}. The structure is modeled in OpenSees~\cite{mckenna2011opensees}, with a total height of 37.19 m and a plan width of 45.73 m. Splice locations for column joints are placed at the first, third, fifth, and seventh stories, each elevated 1.83 m above the beam centerline to accommodate flexural and uplift demands. The foundation is assumed fixed, supported by concrete walls and surrounding soil. Structural responses of interest are the displacement time histories of the nine floors, denoted by $\vect{u}(t) = \{u_1(t), \dots, u_9(t)\}$. Key structural properties are summarized in Table~\ref{Tab_9SAC}, and a structural model is shown in Figure~\ref{Fig_SAC9}. The fundamental period of the system is $T = 2.27$ s.

The excitation dataset consists of 1,000 recorded ground motions selected from the PEER NGA-West2 database~\cite{ancheta2014nga}. Selection criteria include: magnitude $5.0 < M < 7.0$, Joyner–Boore distance $15 < R_{\text{JB}} < 50$ km, average shear wave velocity $180 < V_{S30} < 1800$ m/s, and uniform time step $\Delta t = 0.005$ s. All motions are truncated to a 20-second duration, resulting in 4,001 time steps per sample.

\begin{table}[H]
  \caption{\textbf{Structural parameters of the nine-story steel moment-resisting frame.}}
  \label{Tab_9SAC}
  \centering
  \begin{tabular}{c c}
    \toprule
    Parameter & Value \\
    \midrule
    Damping ratio, \% & 3 \\
    Elastic modulus, GPa & 200 \\
    Yield strength (beam), MPa & 248 \\
    Yield strength (column), MPa & 345 \\
    Strain hardening ratio (beam) & 0.01 \\
    Strain hardening ratio (column) & 0.01 \\
    \bottomrule
  \end{tabular}
\end{table}

\subsubsection{Prediction performance and discussion}

To construct the physics-based intermediate representation $\mathcal{H}_p$, an ELS is derived using pushover analysis. A triangular lateral load pattern is applied to the nonlinear structural model to obtain the base shear–roof displacement curve. The initial elastic stiffness is estimated from the slope of the linear portion of this curve. Story-level stiffness values are then computed by dividing the lateral force at each floor by the corresponding interstory drift, which are used to define an effective stiffness matrix $\mathbf{K}_{\mathrm{eq}}$ in Eq.~\eqref{Eq:ELS}.

Both the C-PhysFNO and baseline FNO models are trained using 600 samples, with 100 held out for validation and 300 used for testing, including a total training time of around 60 minutes. Both models employ 8 Fourier layers, 40 retained spectral modes, and 128 feature dimensions. Lifting and projection layers follow the structure described in Section~\ref{MDOF_BW:results}.

\begin{figure}[H]
  \centering
  \includegraphics[scale=0.67] {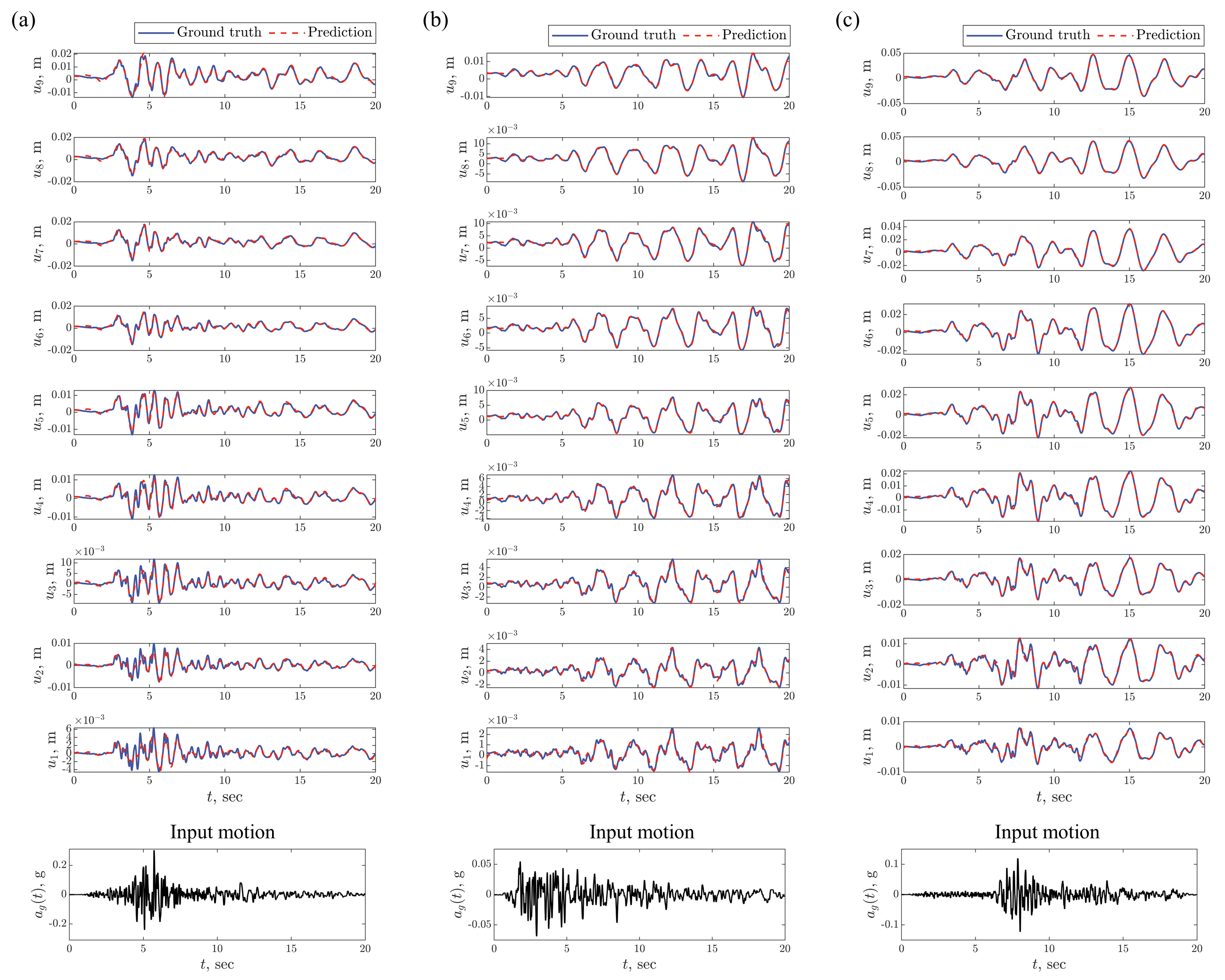}
  \caption{\textbf{Predicted story displacement trajectories using C-PhysFNO for the nine-story steel moment-resisting frame.} Each panel shows the response to one test ground motion: (a) input motion 1 and corresponding prediction, (b) motion 2, and (c) motion 3.}
  \label{Fig_pred_SAC9}
\end{figure}

\begin{figure}[H]
  \centering
  \includegraphics[scale=0.40] {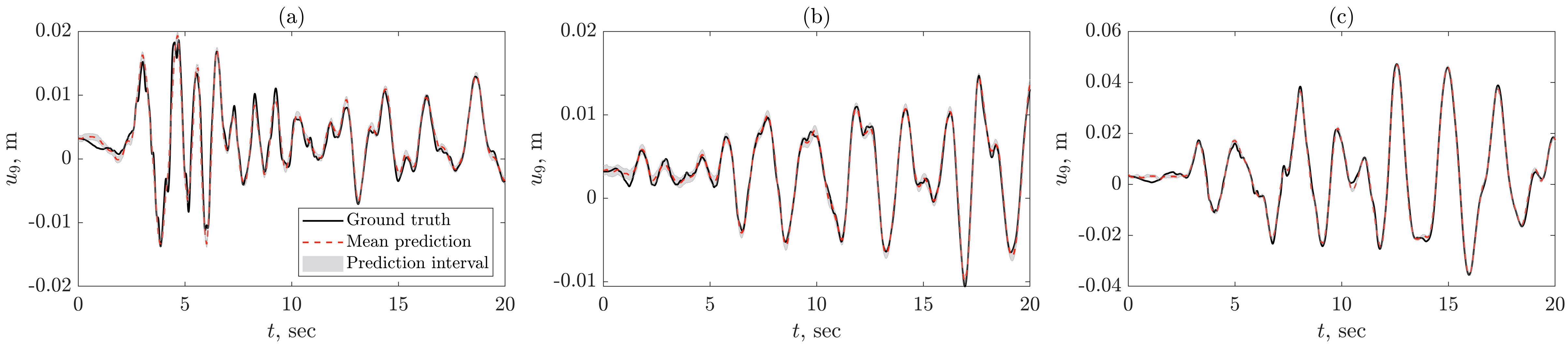}
  \caption{\textbf{Linear regression-based refinement of C-PhysFNO predictions for the nine-story steel moment-resisting frame.} Each panel shows the top-story response for the test ground motions in Figure~\ref{Fig_pred_SAC9}. The shaded region represents the prediction interval defined by the posterior mean $\pm$ one standard deviation.}
  \label{Fig_Post_SAC9}
\end{figure}

\begin{figure}[H]
  \centering
  \includegraphics[scale=0.50] {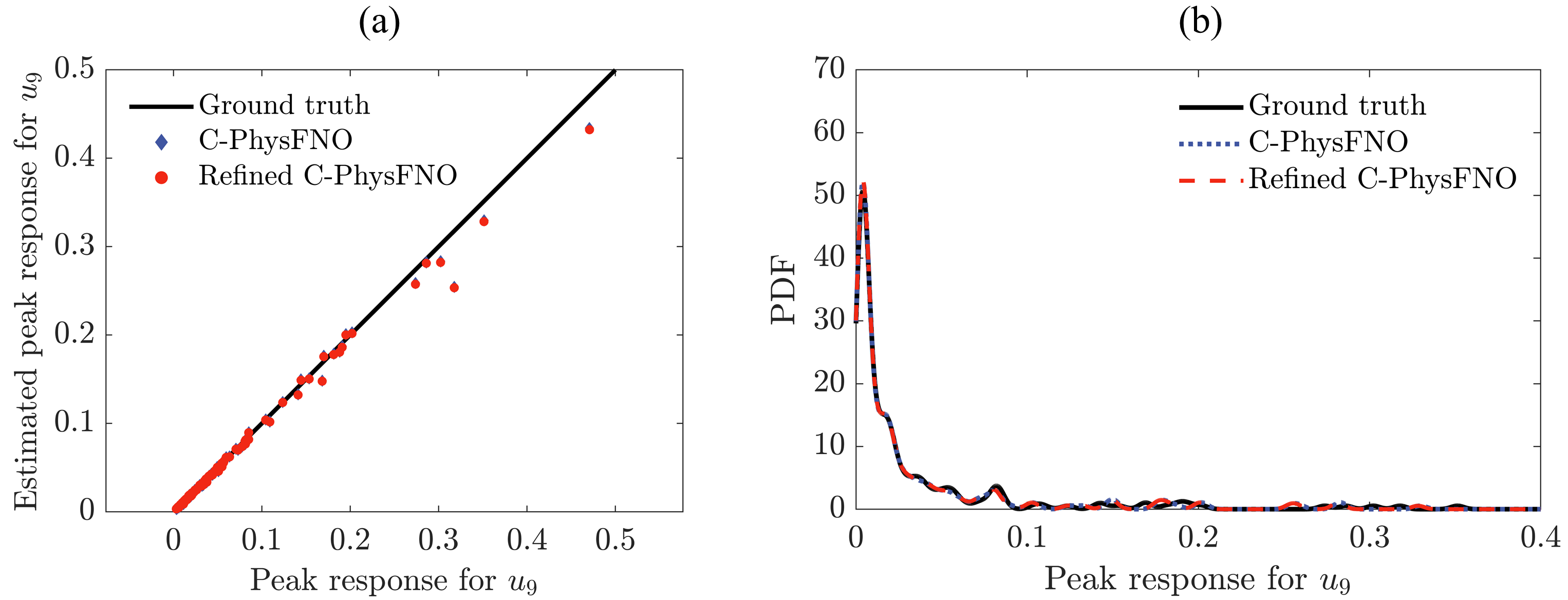}
  \caption{\textbf{Top-story peak displacement prediction for the nine-story steel moment-resisting frame: (a) scatter plot of predicted versus ground truth peak values across the test set, and (b) estimated probability distribution of peak responses.} Both C-PhysFNO and its refinement provide accurate predictions with $R^2$ values of 0.9920 and 0.9914, respectively.}
  \label{Fig_SAC9_Peak}
\end{figure}

Figure~\ref{Fig_pred_SAC9} illustrates the predicted story-level displacement trajectories produced by C-PhysFNO for three distinct test ground motions. In each panel, the input motion and the corresponding predicted response across all stories are shown. The model demonstrates high accuracy in capturing temporal patterns of nonlinear structural responses. Corresponding refined predictions for top story displacements are shown in the Figure~\ref{Fig_Post_SAC9}, which offers smoother trajectories and quantify uncertainty through posterior intervals. Table~\ref{Tab_result_SAC9} summarizes the RMSE and relative $\mathcal{L}^2$ errors across all nine stories, averaged over 10 independent training runs. Across all degrees of freedom, C-PhysFNO consistently achieves lower error than the baseline FNO, confirming the benefit of incorporating physics-based intermediate representations for trajectory-level seismic response prediction under recorded ground motion inputs. In addition, Figure~\ref{Fig_SAC9_Peak} further shows that both C-PhysFNO and its refinement accurately capture peak displacement values, aligning closely with the ground truth across the test set.

From a computational perspective, the total offline cost of building the C-PhysFNO model includes both the data generation and model training phases. In this example, generating 600 full nonlinear simulations required approximately 850 minutes on an Intel i9-13900H CPU, while training the neural operator took an additional 60 minutes on the GPU. By contrast, evaluating the same 300 testing samples using full nonlinear analysis would have incurred over 430 minutes of computation time, whereas the trained C-PhysFNO model completes inference for all test cases in under 5 seconds. This highlights a favorable trade-off: while the initial offline cost is nontrivial, the proposed approach enables significant acceleration in downstream evaluations—particularly advantageous for tasks requiring repeated simulations such as uncertainty quantification, design optimization, or real-time decision support. Such computational efficiency positions the proposed method as a viable surrogate for forward modeling in many-query settings. However, for limited-scope analyses involving only a small number of forward simulations, conventional numerical solvers may remain preferable. These practical considerations reinforce the intended use case of the proposed framework and complement earlier discussions on predictive accuracy and operator generalization.

\begin{table}[H]
  \caption{\textbf{Prediction error metrics for the nine-story steel moment-resisting frame across all neural operator models.}}
  \label{Tab_result_SAC9}
  \centering
  \resizebox{\textwidth}{!}{
  \begin{tabular}{l | l | c c c c c c c c c}
    \toprule
    Model & Error metric & Story 1 & Story 2 & Story 3 & Story 4 & Story 5 & Story 6 & Story 7 & Story 8 & Story 9 \\
    \midrule
    \multirow{2}{*}{FNO} 
    & RMSE (m) & 0.0011  &  0.0017 &    0.0024  &  0.0031  &  0.0035  &  0.0040  &  0.0045  &  0.0051  &  0.0056 \\
    & Relative $\mathcal{L}^2$  & 0.2629  &  0.2481  &  0.2458 &   0.2481  &  0.2381  &  0.2345 &   0.2318  &  0.2296  &  0.2323 \\
    \midrule
    \multirow{2}{*}{C-PhysFNO} 
    & RMSE (m) & 0.0009 &   0.0015 &   0.0020 &   0.0024 &  0.0028 &   0.0032 &  0.0036 &   0.0041 &  0.0046 \\
    & Relative $\mathcal{L}^2$  & 0.2055  &  0.1978 &   0.1889  &  0.1795 &  0.1753  &  0.1718 &  0.1679  &  0.1693 &  0.1749  \\
    \midrule
    \multirow{2}{*}{\shortstack{Refined\\C-PhysFNO}} 
    & RMSE (m) & 0.0009 &   0.0015 &   0.0020 &   0.0024 &  0.0028 &   0.0032 &  0.0035 &   0.0041 &  0.0045 \\
    & Relative $\mathcal{L}^2$ & 0.2054  &  0.1976 &   0.1887  &  0.1794 &  0.1752  &  0.1716 &  0.1678  &  0.1692 &  0.1748  \\
    \bottomrule
  \end{tabular}}
\end{table}

\section{Conclusions} \label{Conclusion}

This study presents a composite physics-neural operator (C-PhysFNO) method for accurate and efficient prediction of nonlinear structural response trajectories under seismic excitation. The proposed approach introduces a composite operator learning architecture that integrates simplified physics-based models, such as equivalent linear systems, modal reductions, and relaxed solvers, as intermediate representations. These coarse-grained models embed essential structural dynamics while reducing the functional complexity of the learning task, thereby enabling effective function-to-function regression from ground motion inputs to full nonlinear structural responses. In addition, this study introduces a linear regression–based postprocessing step to refine predictions and quantify uncertainty without modifying the neural operator architecture.

Extensive numerical experiments are conducted on three representative structural systems: a five-story nonlinear building with Bouc–Wen hysteretic behavior, a prestressed concrete bridge subjected to synthetic motions, and a nine-story steel moment-resisting frame subjected to recorded ground motions. In all cases, the C-PhysFNO approaches consistently outperform standard neural operator baselines in terms of prediction accuracy and generalization. The linear regression-based postprocessing further refines predictions and provides a preliminary quantification of prediction uncertainty. These improvements are particularly pronounced in data-limited settings and under broadband seismic excitations.

Several promising directions remain for future research. First, the integration of adaptive sampling and active learning strategies~\cite{kim2024adaptive,zhou2025multi} into the C-PhysFNO framework could further improve data efficiency by prioritizing regions of high predictive uncertainty or critical peak response metrics, especially for probabilistic risk assessment applications. Second, while the present framework assumes deterministic structural parameters, extending it to incorporate epistemic uncertainties in material properties, geometric configurations, and boundary conditions would enable broader applications in structural reliability and design optimization~\cite{celik2010seismic,kim2024active}. Finally, although the current framework leverages postprocessing via linear regression to quantify predictive uncertainty, directly integrating uncertainty modeling into the neural operator, such as through Bayesian neural operators~\cite{magnani2022approximate,bulte2025probabilistic} could further advance trajectory-level uncertainty quantification.

\section*{Acknowledgments}
This research was supported by the Basic Science Research Program through the National Research Foundation of Korea (NRF) funded by the Ministry of Education (RS-2024-00407901). This work was also supported by the faculty research fund of Sejong University in 2025. The authors also thank the two anonymous reviewers for their thoughtful and constructive comments, which greatly improved the quality of this work.

\section*{Data availability}
The source codes are available for download at \url{https://github.com/Jungh0Kim/C-PhysFNO} upon publication.

\bibliography{NO_seismic}

\end{document}